\documentclass[aps,prl,reprint,superscriptaddress,nofootinbib,longbibliography,showkeys,floatfix]{revtex4-2}

\usepackage[utf8]{inputenc}
\usepackage[T1]{fontenc}
\usepackage{amsmath,amssymb,bm,mathtools}
\usepackage{graphicx}
\usepackage{dcolumn}
\usepackage{booktabs}
\usepackage{multirow}
\usepackage{tabularx}
\usepackage{xcolor}
\usepackage{url}
\usepackage[colorlinks=true,citecolor=blue,linkcolor=blue,urlcolor=blue]{hyperref}

\begin{document}

\title{String Model Predicts Relaxation in Sheared Glass-Forming Liquids}

\author{Zi-Long Wang}
\affiliation{State Key Laboratory of Polymer Science and Technology, Changchun Institute of Applied Chemistry, Chinese Academy of Sciences, Changchun 130022, P. R. China}
\affiliation{School of Applied Chemistry and Engineering, University of Science and Technology of China, Hefei 230026, P. R. China}

\author{Qi-Lu Yuan}
\affiliation{State Key Laboratory of Polymer Science and Technology, Changchun Institute of Applied Chemistry, Chinese Academy of Sciences, Changchun 130022, P. R. China}
\affiliation{School of Applied Chemistry and Engineering, University of Science and Technology of China, Hefei 230026, P. R. China}

\author{Yun-Jiang Wang}
\affiliation{State Key Laboratory of Nonlinear Mechanics, Institute of Mechanics, Chinese Academy of Sciences, Beijing 100190, P. R. China}
\affiliation{School of Engineering Science, University of Chinese Academy of Sciences, Beijing 100049, P. R. China}

\author{Jack F. Douglas}
\email{jack.douglas@nist.gov}
\affiliation{Materials Science and Engineering Division, National Institute of Standards and Technology, Gaithersburg, Maryland 20899, United States}

\author{Matteo Baggioli}
\email{b.matteo@sjtu.edu.cn}
\affiliation{Wilczek Quantum Center, School of Physics and Astronomy, Shanghai Jiao Tong University, Shanghai 200240, P. R. China}
\affiliation{Shanghai Research Center for Quantum Sciences, Shanghai 201315, P. R. China}

\author{Zhao-Yan Sun}
\email{zysun@ciac.ac.cn}
\affiliation{State Key Laboratory of Polymer Science and Technology, Changchun Institute of Applied Chemistry, Chinese Academy of Sciences, Changchun 130022, P. R. China}
\affiliation{School of Applied Chemistry and Engineering, University of Science and Technology of China, Hefei 230026, P. R. China}
\affiliation{Jilin Provincial International Cooperation Key Laboratory for Polymer Processing Physics, Changchun 130022, P. R. China}

\author{Wen-Sheng Xu}
\email{wsxu@nju.edu.cn}
\affiliation{MOE Key Laboratory of High Performance Polymer Materials and Technology, State Key Laboratory of Coordination Chemistry, School of Chemistry, Nanjing University, Nanjing 210023, P. R. China}

\date{\today}

\begin{abstract}
Understanding how structural relaxation evolves from equilibrium to nonequilibrium conditions remains a central problem in glass physics. Using simulations of model glass formers under steady shear, we show that the String Model, which links stringlike cooperative rearrangements to relaxation in equilibrium glass-forming liquids, predicts the structural relaxation time across the full range of temperatures and shear rates investigated without additional nonequilibrium fitting parameters, when combined with a shear-dependent effective temperature that is captured by a generalized fluctuation-dissipation relation. These results indicate that equilibrium and nonequilibrium relaxation are governed by the same underlying cooperative mechanism, but they occur under different effective thermodynamic conditions under steady shear. More broadly, our study provides a unified microscopic description of thermal and mechanically driven dynamics in glass-forming liquids.
\end{abstract}

\maketitle

\color{blue} \textit{Introduction} \color{black} -- The nature of glass formation remains one of the central unresolved problems in condensed matter physics \cite{doi:10.1126/science.267.5204.1615.f}. Upon cooling molecular liquids or compressing colloidal suspensions, viscosity and structural relaxation times increase by many orders of magnitude, whereas static structural measures exhibit only modest changes and remain largely liquid-like \cite{Supercooled_1996_100_13200}. Explaining this dramatic slowdown in the absence of obvious structural signatures lies at the heart of the glass problem \cite{Theoretical_2011_83_587}. The task becomes significantly more complex under nonequilibrium conditions, such as physical aging \cite{Narayanaswamys_2015_143_114507, Physical_1978__, Predicting_2022_8_eabl9809} or externally driven states \cite{Fluctuation_2000_63_012503, Dynamics_1998_58_3515, Anisotropic_2009_102_016001, Structural_2018_115_87, Predictive_2020_6_eaaz0777}.

Over the past few decades, growing evidence has linked the dramatic slowdown of glass-forming liquids to the emergence of collective particle motion \cite{Weeks_2000_287_627, Kob_1997_79_2827}. Among the various manifestations of dynamic heterogeneity \cite{Berthier_2011_Dynamical}, stringlike cooperative motion \cite{Donati_1998_80_2338} has emerged as one of the most ubiquitous signatures of glass formation. Building on this observation, the String Model \cite{String_2014_140_204509}---a microscopic realization of the Adam-Gibbs theory \cite{Temperature_1965_43_139}---relates structural relaxation directly to the extent of stringlike cooperative rearrangements and quantitatively describes relaxation in a wide range of glass-forming materials, including polymers and metallic liquids \cite{Polymer_2021_54_3001}. Systematic studies have further connected stringlike cooperative motion to activation barriers, configurational entropy, and emergent rigidity \cite{Scaling_1994_73_1376, Testing_1995_51_4626, Role_2015_142_164506, Parallel_2023_56_4929, Quantitative_2015_112_2966}. Collectively, these findings establish stringlike cooperative motion as a unifying microscopic framework for understanding equilibrium glassy dynamics and the origin of characteristic glass anomalies, including the boson peak \cite{tanakaNatPhys, tanakaPRR}.

Whether equilibrium and nonequilibrium relaxation are governed by the same microscopic mechanism remains unclear. Mechanical deformation can accelerate structural relaxation by many orders of magnitude, effectively reversing the dynamical slowdown associated with glass formation \cite{Berthier_2002_Shearing, Prasad_2007_Confocal}. Despite decades of study \cite{Fuchs_2002_Theory, Falk_1998_Dynamics}, no predictive framework has successfully connected the cooperative motion that controls equilibrium relaxation to the strongly accelerated dynamics observed under external driving \cite{Brader_2010_Nonlinear}. This raises a fundamental question: can the same microscopic mechanism that explains equilibrium relaxation also explain shear-driven relaxation? 

In this Letter, we study relaxation in glass-forming liquids under steady shear, a typical nonequilibrium state having direct relevance for materials processing. We show that equilibrium and shear-driven relaxation originate from the same underlying cooperative mechanism. However, the suppression of stringlike cooperative motion alone is insufficient to account for the accelerated dynamics induced by external driving. To resolve this discrepancy, we combine collective motion with a nonequilibrium effective temperature captured by a generalized fluctuation-dissipation relation (FDR) framework \cite{Fluctuation_2000_63_012503}. Together, these two ingredients quantitatively predict the structural relaxation time across a broad range of temperatures and shear rates without additional nonequilibrium fitting parameters. More broadly, our results establish a unified microscopic description of equilibrium and nonequilibrium relaxation, in which external driving acts primarily by modifying the effective thermodynamic conditions governing activated dynamics.

\color{blue}\textit{Results} \color{black} -- We study three representative glass-forming liquids: the Kob--Andersen binary Lennard--Jones model (KABLJ) \cite{Scaling_1994_73_1376, Testing_1995_51_4626} and a coarse-grained bead-spring polymer melt \cite{Dynamics_1990_92_5057, Molecular_1986_33_3628} under constant-volume (Poly-V) and constant-pressure (Poly-P) conditions. All quantities and results are expressed in standard reduced units; e.g., length, time, temperature, and pressure are measured in units of $\sigma$, $\tau=\sqrt{m\sigma^2/\varepsilon}$, $\varepsilon/k_{\mathrm B}$, and $\varepsilon/\sigma^3$, respectively, where $\sigma$ and $\varepsilon$ are the length and energy parameters of the Lennard--Jones (LJ) potential, $m$ is the particle mass, and $k_{\mathrm B}$ is Boltzmann's constant. More simulation details are given in the Supplemental Material (SM).

\begin{figure}[htb!]
	\centering
	\includegraphics[angle=0, width=0.48\textwidth]{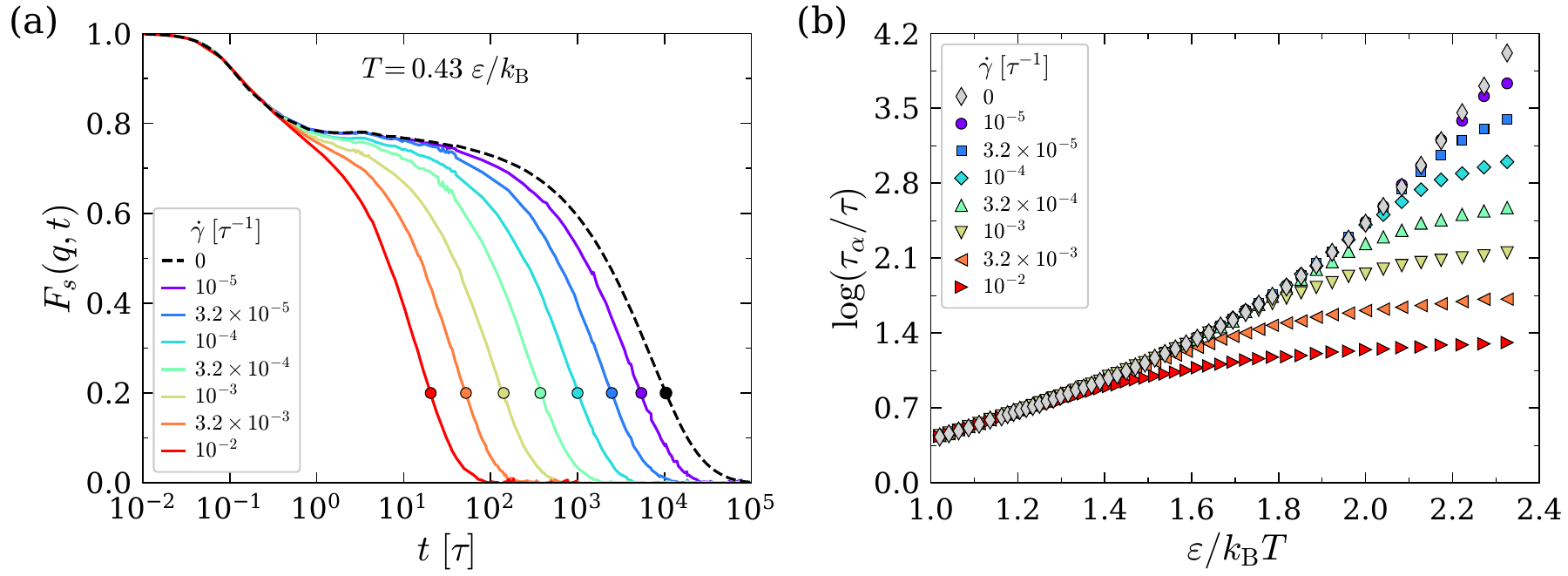}
	\caption{Accelerated structural relaxation under steady shear. (a) Self-intermediate scattering function, $F_s(q,t)$, for the KABLJ system at a temperature of $T=0.43\, \varepsilon/k_{\mathrm B}$ and different shear rates $\dot{\gamma}$. Colored symbols indicate the condition $F_s(q,t)=0.2$, which is used operationally to define the structural relaxation time, $\tau_\alpha$. (b) Logarithm of the structural relaxation time, $\log(\tau_\alpha)$, as a function of $\varepsilon/k_{\mathrm B}T$ for different $\dot{\gamma}$.}
	\label{Fig1}
\end{figure}

We first investigate how steady shear modifies structural relaxation in glass-forming liquids, taking the KABLJ system as a benchmark system. As shown in Fig.~\ref{Fig1}(a), increasing the shear rate $\dot{\gamma}$ at a fixed temperature ($T$) progressively shifts the decay of the self-intermediate scattering function $F_s(q, t)$ to shorter times, indicating a dramatic enhancement of relaxation dynamics. This shear-induced acceleration of structural relaxation is consistent with previous simulations \cite{Fluctuation_2000_63_012503, Dynamics_1998_58_3515, Anisotropic_2009_102_016001, Structural_2018_115_87, Predictive_2020_6_eaaz0777} and experiments \cite{Three_2007_99_028301}. We extract the structural relaxation time $\tau_\alpha$ based on the definition of $F_s(q, \tau_\alpha) = 0.2$. Figure~\ref{Fig1}(b) shows the $T$ dependence of $\tau_\alpha$ for different shear rates. As the shear rate increases at fixed $T$, $\tau_\alpha$ consistently decreases, revealing the profound acceleration of structural relaxation induced by steady shear. Consistent results are found also for the polymer systems and are presented in the SM.

\begin{figure}[htb!]
	\centering
	\includegraphics[angle=0, width=0.48\textwidth]{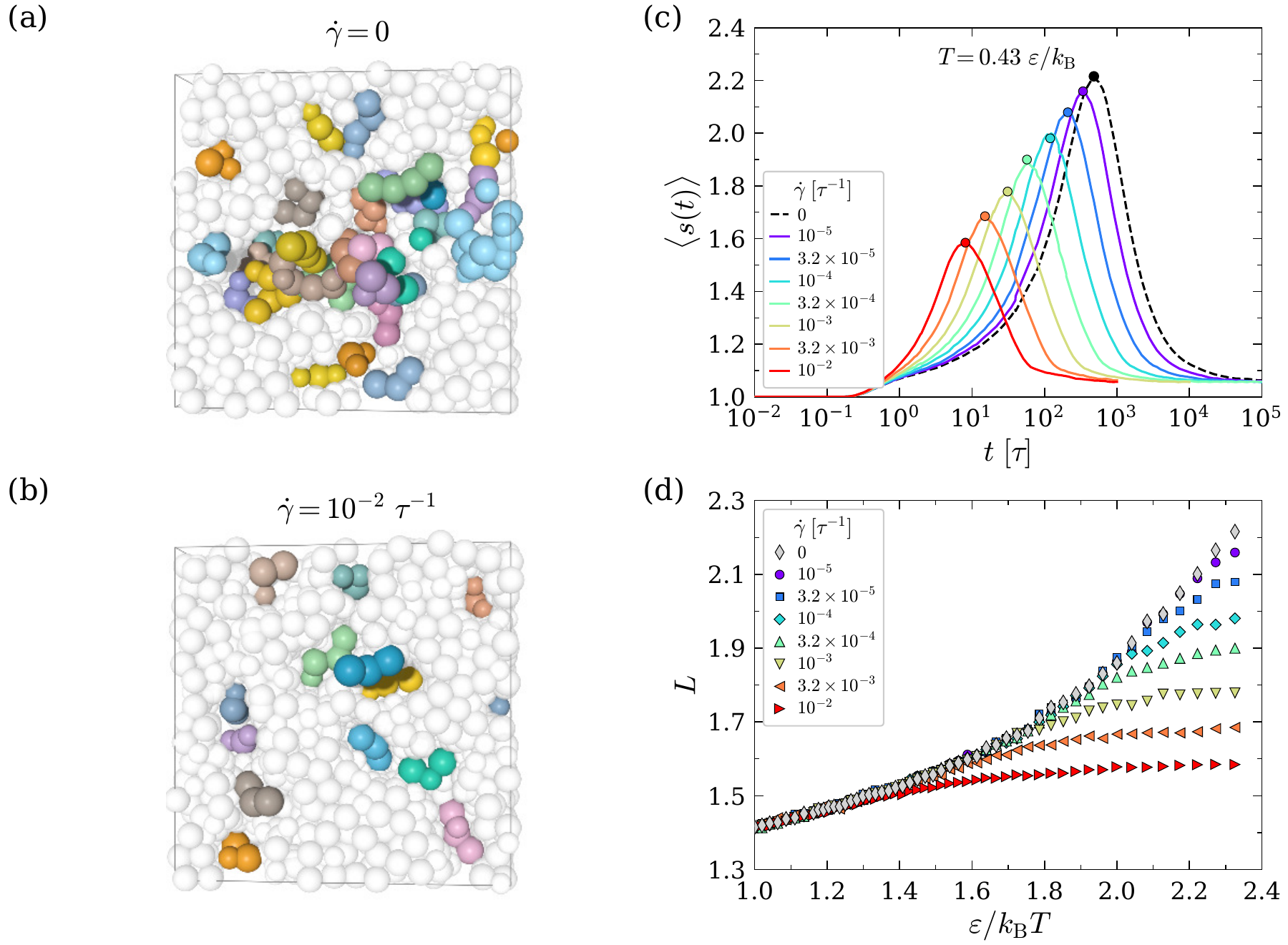}
	\caption{Shear suppresses the collective motion underlying glassy relaxation. (a,b) Representative snapshots of stringlike cooperative motion in the KABLJ system at $T=0.43\,\varepsilon/k_{\mathrm B}$ for shear rates $\dot{\gamma}=0$ and $\dot{\gamma}=10^{-2} \, \tau^{-1}$, respectively. Particles belonging to the same string are shown in the same color. (c) Number-averaged string length, $\langle s(t)\rangle$, as a function of time for different shear rates at $T=0.43\,\varepsilon/k_{\mathrm B}$. (d) Characteristic string length $L$ as a function of $\varepsilon/k_{\mathrm B}T$ for different $\dot{\gamma}$.}
	\label{Fig2}
\end{figure}

We investigate the microscopic origin of shear-induced relaxation by examining the evolution of stringlike cooperative motion with increasing shear rate. Representative snapshots of the KABLJ system in equilibrium and under steady shear are shown in Fig.~\ref{Fig2}(a,b), where particles belonging to the same string are colored identically. The nonequilibrium state exhibits a marked reduction in both the size and prevalence of cooperative strings, indicating a strong suppression of collective motion by external driving. Previous studies have shown that the relaxation and transport properties of glass-forming liquids do not become appreciably anisotropic under steady shear~\cite{Dynamics_1998_58_3515, Anisotropic_2009_102_016001, Structural_2018_115_87}, a finding corroborated by our own simulations (see the SM). Consequently, our treatment relies on an isotropic approximation of the dynamics.

A more quantitative picture emerges from Fig.~\ref{Fig2}(c), which shows the number-averaged string length, $\langle s(t)\rangle$, as a function of time for different $\dot{\gamma}$. Increasing shear rate strongly suppresses the population of cooperative strings. Moreover, both the height and position of the peak in $\langle s(t)\rangle$ decrease with increasing $\dot{\gamma}$, indicating that steady shear suppresses not only the characteristic size of cooperative rearrangements but also the timescale over which they develop. To characterize this behavior, we define the characteristic string length as $L \equiv \langle s(t_L)\rangle$, where $t_L$ is the time at which $\langle s(t)\rangle$ attains its maximum. Figure~\ref{Fig2}(d) shows the evolution of $L$ with shear rate. Below a characteristic temperature, $L$ decreases markedly with increasing shear, revealing a strong suppression of collective motion under nonequilibrium conditions. Evidently, the evolution of $L$ at least qualitatively follows the structural relaxation time $\tau_\alpha$ shown in Fig.~\ref{Fig1}(b), suggesting a direct connection between the reduction of cooperative motion and the acceleration of relaxation dynamics. Similar behavior is observed in the polymer systems (see the SM).

\begin{figure*}[htb!]
	\centering
	\includegraphics[angle=0, width=0.9\textwidth]{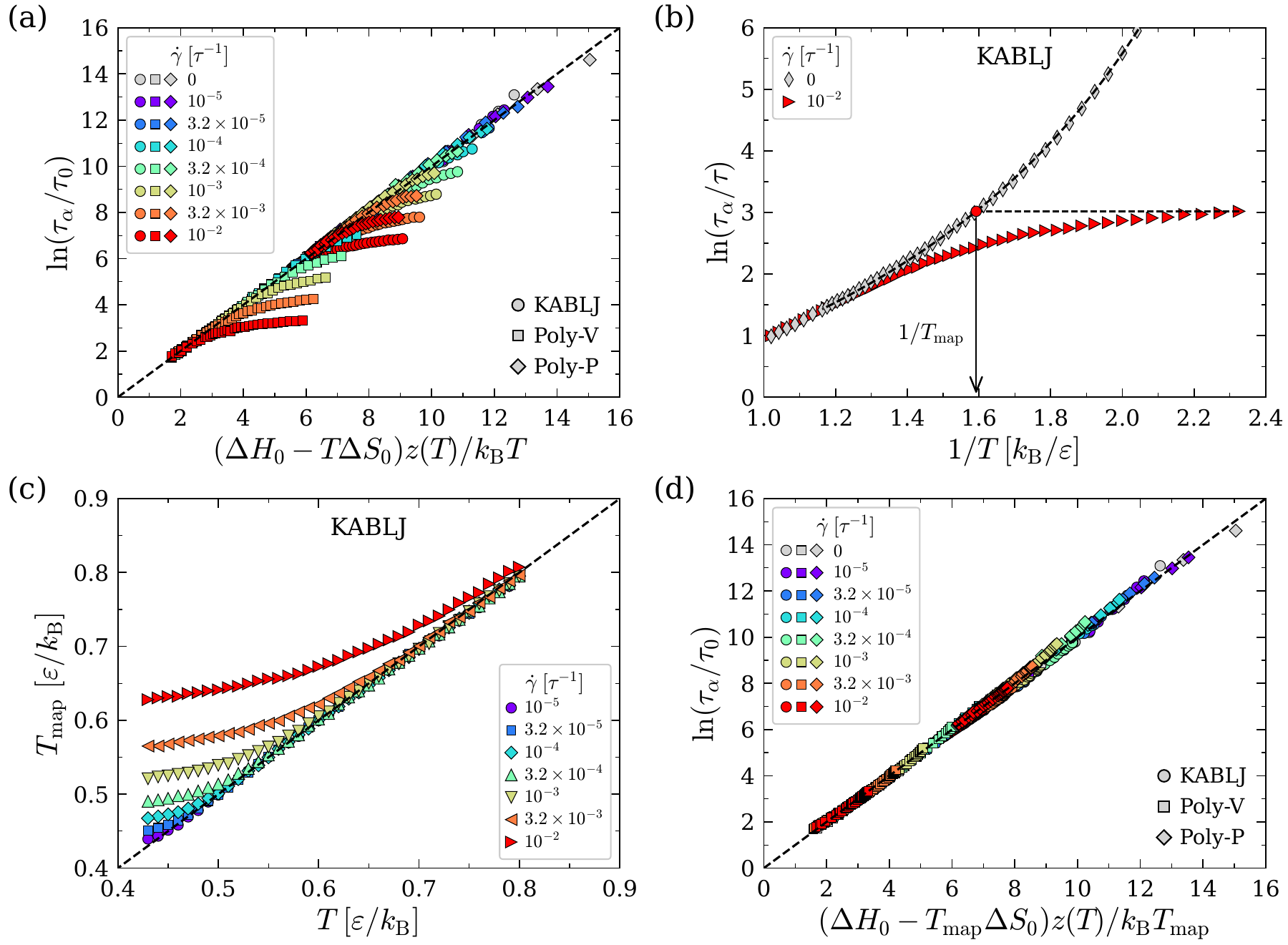}
	\caption{Predicting nonequilibrium relaxation. (a) Logarithm of the structural relaxation time, $\ln(\tau_{\alpha}/\tau_{0})$, is plotted against $(\Delta H_{0}-T\Delta S_{0})z(T)/k_{\mathrm{B}}T$ for three glass-forming systems over a range of shear rates $\dot{\gamma}$. The parameters $\tau_{0}$, $\Delta H_{0}$, and $\Delta S_{0}$ are obtained from equilibrium simulations and are reported in Table~\ref{Table1}. The black dashed line denotes the prediction of the String Model, corresponding to a one-to-one relation between the two quantities. (b) Schematic illustration of the effective temperature $T_{\mathrm{map}}$. (c) Effective temperature $T_{\mathrm{map}}$ as a function of the bath temperature $T$ at different shear rates for the KABLJ system. (d) $\ln(\tau_{\alpha}/\tau_{0})$ as a function of $(\Delta H_{0}-T_{\mathrm{map}}\Delta S_{0})z(T)/k_{\mathrm{B}}T_{\mathrm{map}}$ for all three systems and a wide range of shear rates. The data collapse onto the predicted linear relation, shown by the black dashed line.}
	\label{Fig3}
\end{figure*}

\begin{table}[htbp]
	\centering
	\caption{Basic properties of glass formation for the three systems at equilibrium. $T_A$ is the onset temperature of glass formation. $\Delta H_0$ and $\Delta S_0$ are the enthalpy and entropy of high temperature activation. $\tau_A$ is the relaxation time at $T_A$. $\tau_{0}$ is the prefactor in the String Model.}
	\label{Table1}
	\begin{ruledtabular}
	\begin{tabular}{lccccc}
	System & 
	$T_A \ [\varepsilon/k_{\text{B}}]$ & 
	$\Delta H_0 \ [\varepsilon]$ & 
	$\Delta S_0 \ [k_{\text{B}}]$ & 
	$\tau_A \ [\tau]$ & 
	$\tau_0 \ [\tau]$ \\
	\colrule
	KABLJ & 0.80 & 2.70 & -2.16 & 5.41 & $2.14 \times 10^{-2}$ \\
	Poly-V & 0.75 & 2.09 & 1.07 & 3.66 & $6.60 \times 10^{-1}$ \\
	Poly-P & 0.65 & 2.03 & -3.04 & 3.48 & $7.27 \times 10^{-3}$ \\
	\end{tabular}
	\end{ruledtabular}
\end{table}

The striking similarity between the shear dependence of $L$ in Fig.~\ref{Fig2}(d) and that of $\tau_\alpha$ in Fig.~\ref{Fig1}(b) suggests that stringlike cooperative motion may provide the microscopic mechanism underlying relaxation in both equilibrium and nonequilibrium glass-forming liquids. To test this possibility, we turn to the String Model of glass formation \cite{String_2014_140_204509, Communication_2014_141_141102, Polymer_2021_54_3001}. Rooted in transition state theory \cite{Theory_1941_28_301, Book_Eyring} and broadly consistent with entropy-based theories of glass formation \cite{Temperature_1965_43_139}, the String Model posits that, below the onset temperature $T_A$ where the dynamics starts to deviate from Arrhenius behavior, the activation free energy for structural relaxation grows in proportion to the size of cooperative strings. Specifically, $\Delta G(T)=\Delta G_0 z(T)$, where $\Delta G_0=\Delta H_0-T_A\Delta S_0$ is the activation free energy at $T_A$, and $z(T)=L(T)/L_A$ with $L_A \equiv L(T_A)$. This assumption leads directly to the String Model expression for $\tau_{\alpha}$,
\begin{equation} 
	\label{theory} 
	\tau_{\alpha} = \tau_{0} \exp\left[ \frac{\Delta H_0 - T \Delta S_0}{k_{\mathrm{B}} T} z(T) \right].
\end{equation}
The prefactor $\tau_0$ is fixed by the relaxation time at $T_A$, $\tau_A=\tau_\alpha(T_A)$, while the activation enthalpy $\Delta H_0$ is independently determined from the Arrhenius regime at high temperatures above $T_A$. Consequently, $\Delta S_0$ is the only fitting parameter. As shown in the SM, Eq.~\eqref{theory} quantitatively captures the equilibrium dynamics of all three systems considered here, with the relevant parameters summarized in Table~\ref{Table1}.

As a first minimal extension of the String Model, we assume that the sole effect of steady shear is to suppress the characteristic string length $L$ entering the parameter $z(T)$ in Eq.~\eqref{theory}, while all other quantities are retained at their equilibrium values determined from the equilibrium data. We test this hypothesis in Fig.~\ref{Fig3}(a). Despite capturing the reduction of cooperative motion, the resulting predictions show evident systematic deviations from the simulation data. This demonstrates that the suppression of stringlike cooperative rearrangements alone is insufficient to account for the nonequilibrium acceleration of structural relaxation under steady shear.

Because steady shear drives the system out of thermal equilibrium, the bath temperature $T$ is no longer sufficient to characterize the relaxation dynamics. It is then sensible to introduce an effective temperature for structural relaxation on the timescale used to define $\tau_\alpha$. In this picture, the driven relaxation at this timescale is mapped onto that of an equilibrium system at a higher temperature. Importantly, this choice preserves the equilibrium theoretical framework with minimal modification, requiring no additional nonequilibrium fitting parameters. It also avoids the need for \textit{ad hoc} shear-dependent parametrizations of activation barriers, as commonly employed in phenomenological approaches such as the Eyring model \cite{Viscosity_1936_4_283, Microscopic_2008_41_5908, Probing_2017_114_7952}.

More precisely, we define an effective temperature $T_{\mathrm{map}}$ as the equilibrium temperature for which the equilibrium structural relaxation time equals that of the driven system [Fig.~\ref{Fig3}(b)],
\begin{equation}
	\label{Eq_Tmap}
	\tau_{\alpha}(T_{\mathrm{map}}, \dot{\gamma}=0) = \tau_{\alpha}(T, \dot{\gamma}).
\end{equation}
As shown in Fig.~\ref{Fig3}(c) for the KABLJ system, $T_{\mathrm{map}}$ is consistently larger than the bath temperature $T$ below $T_A$ and increases monotonically with $\dot{\gamma}$. Thus, stronger driving effectively moves the system toward thermodynamic conditions associated with higher equilibrium temperatures.

We next hypothesize that structural relaxation under steady shear remains governed by the String Model, provided that the nonequilibrium driving is incorporated through two quantities: the shear-dependent cooperative motion encoded in $z(T)$ and the effective temperature $T_{\mathrm{map}}$. The relaxation time is therefore predicted to obey
\begin{equation}
	\label{theory2}
	\tau_{\alpha} = \tau_{0} \exp\left[\frac{\Delta H_{0} - T_{\mathrm{map}} \Delta S_{0}}{k_{\mathrm{B}} T_{\mathrm{map}}} z(T)
\right].
\end{equation}
In this description, shear affects relaxation in two ways: it suppresses stringlike cooperative motion, thereby modifying $z(T)$, and it shifts the relaxation dynamics at the bath temperature $T$ to a higher effective temperature, $T_{\mathrm{map}}>T$. Importantly, after fixing the parameters from equilibrium simulations, the relaxation time under nonequilibrium conditions is predicted without introducing any additional fitting parameters.

The prediction of Eq.~\eqref{theory2} is tested in Fig.~\ref{Fig3}(d) over a broad range of temperatures and shear rates for three distinct glass-forming systems. The agreement is remarkable. Without introducing any additional nonequilibrium fitting parameters, the model quantitatively captures the relaxation dynamics across both equilibrium and driven conditions. These results provide strong evidence that stringlike cooperative motion remains the microscopic mechanism governing structural relaxation under steady shear, while the effects of driving are largely captured through $T_{\mathrm{map}}$. To exclude the possibility that this success merely reflects the definition of $T_{\mathrm{map}}$ from $\tau_\alpha$, we show in the SM that $T_{\mathrm{map}}$ alone, without the cooperativity factor $z(T, \dot{\gamma})$, fails to collapse data across different systems. The combination of both ingredients is therefore essential for a quantitative, system-independent description.

Despite the excellent agreement between Eq.~\eqref{theory2} and the simulation data, the mapping temperature \(T_{\mathrm{map}}\) is introduced phenomenologically. We therefore seek a microscopic interpretation in terms of nonequilibrium FDRs. As detailed in the End Matter, we construct an effective-temperature measure from the equilibrium states dynamically sampled by the driven liquid. We find that the resulting averaged temperature \(T_{\mathrm{avg},c}\) coincides with \(T_{\mathrm{map}}\), whereas the conventional effective temperature \(T_{\mathrm{slow},c}\), extracted solely from the slow branch of the FDR, does not reproduce the shear-dependent structural relaxation time. Because \(T_{\mathrm{avg},c}\) is determined independently from the fluctuation--dissipation construction, without using the value of \(\tau_\alpha\), this agreement provides an independent determination of the temperature entering Eq.~\eqref{theory2} and further resolves the apparent circularity associated with the determination of \(T_{\mathrm{map}}\) through \(\tau_\alpha\).

This distinction is physically significant. Previous work has shown that effective disorder temperatures can control the activation of particle rearrangements under driving \cite{PhysRevLett.99.195701}. Here, we find that \(T_{\mathrm{avg},c}\) also provides the thermodynamic scale governing cooperative stringlike motion and, consequently, the structural relaxation time. The failure of \(T_{\mathrm{slow},c}\) suggests that steady shear destroys a clean separation between slow structural relaxation and faster cage-scale dynamics: the imposed advective timescale truncates cooperative rearrangements and mixes dynamical sectors that are more clearly separated near equilibrium. Because \(T_{\mathrm{avg},c}\) incorporates the response across the full relaxation process rather than only its asymptotic slow sector, it naturally captures this mode mixing. The fluctuation--dissipation construction therefore provides a physical basis for \(T_{\mathrm{map}}\) and identifies it as the effective thermodynamic variable required to extend the String Model to nonequilibrium steady states.

\color{blue} \textit{Discussion} \color{black} -- In summary, we have investigated the microscopic origin of structural relaxation in glass-forming liquids under steady shear. Our simulations show that external driving progressively suppresses stringlike cooperative motion and dramatically accelerates relaxation. While the equilibrium String Model fails when applied directly under shear, introducing $T_{\mathrm{map}}$ as the effective control temperature restores its predictive power. The resulting nonequilibrium String Model quantitatively describes the relaxation time across different temperatures, shear rates, and glass-forming systems without introducing any additional nonequilibrium fitting parameters. 

These findings identify stringlike cooperative motion as the common microscopic mechanism governing relaxation in both equilibrium and driven glass-forming liquids, while revealing the effective temperature as the key quantity linking thermal and mechanical activation. More broadly, our results suggest that external driving acts primarily by modifying the effective thermodynamic conditions that control collective activated motion, providing a unified microscopic description of relaxation in equilibrium and nonequilibrium glass-forming liquids. Whether this effective-temperature description can be extended to other nonequilibrium conditions, including aging and oscillatory driving, remains an important question for future work.

\color{blue}\textit{Acknowledgments} \color{black} -- This work was supported by the National Natural Science Foundation of China (Nos. 22573102 and 52293471) and the Jilin Provincial Science and Technology Development Program (Nos. SKL202602006JC and SKL202502004JC). Z.-Y.S. is grateful for the essential support of Jilin Provincial International Cooperation Key Laboratory for Polymer Processing Physics. M.B. acknowledges the support of the Foreign Young Scholars Research Fund Project (Grant No. 22Z033100604) and the sponsorship from the Yangyang Development Fund.

\section*{Data Availability}
The data used to generate the figures presented in this manuscript are available from the Open Science Framework: \url{https://osf.io/9fvdw}.\\

\bibliographystyle{apsrev4-2}
\bibliography{refs}

\section{End Matter}\label{end}
{\it Appendix A: Effective Temperature from Nonequilibrium Fluctuation–Dissipation Relation} -- To obtain a more grounded definition of $T_{\mathrm{map}}$, we begin from the concept of an effective temperature $T_{\mathrm{eff}}$, introduced through a nonequilibrium generalization of the fluctuation–dissipation relation (FDR) \cite{Energy_1997_55_3898},
\begin{equation}\label{ee}
R(t) = -\frac{1}{k_{\mathrm{B}} T_{\mathrm{eff}}(C)} \frac{dC(t)}{dt},
\end{equation}
where $R(t)$ and $C(t)$ denote the response and correlation functions, respectively. In equilibrium, the standard FDR is recovered with $T_{\mathrm{eff}}(C)=T$ at all times.

Introducing the susceptibility $\chi(t)=\int_0^{t} R(t')\,dt'$, \eqref{ee} can be rewritten as
\begin{equation}
\frac{d\chi}{dC} = -\frac{1}{k_{\mathrm{B}} T_{\mathrm{eff}}(C)}.
\label{ee2}
\end{equation}
In this form, a parametric plot of $\chi$ versus $C$ becomes piecewise linear, with each linear sector defining a distinct effective temperature.

In supercooled liquids driven out of equilibrium, two regimes are typically observed. Fast degrees of freedom obey the equilibrium FDR, yielding a slope $-1/(k_{\mathrm{B}}T)$, while slow structural modes define a second linear sector associated with a higher effective temperature $T_{\mathrm{slow}} > T$ \cite{Fluctuation_2000_63_012503,10.1063/1.1460862}. The former reflects short-time vibrational dynamics that remain equilibrated with the bath, whereas the latter encodes the long-time configurational response.

An alternative characterization is obtained by averaging over the full FDR curve \cite{PhysRevLett.101.245701,Langer_2000},
\begin{equation}
\frac{1}{k_{\mathrm{B}} T_{\mathrm{avg}}} = \chi(t \rightarrow \infty)= \int_0^1 \frac{dC'}{k_{\mathrm{B}} T_{\mathrm{eff}}(C')},
\label{eeav}
\end{equation}
which corresponds to a uniform-in-\(C\) coarse-graining of the inverse effective temperature over the full relaxation process.

$T_{\mathrm{avg}}$ is often referred to as the \emph{effective disorder temperature} \cite{Steady_2007_76_056107,nonequilibrium_2009_80_031132}. It has been widely used in driven amorphous materials and differs from $T_{\mathrm{slow}}$ defined above. Rather than probing only the asymptotic long-time sector, it provides a single measure of the full configurational response of the system.

Importantly, neither $T_{\mathrm{slow}}$ nor $T_{\mathrm{avg}}$ coincides with the phenomenological mapping temperature $T_{\mathrm{map}}$ (see the SM), and both fail to reproduce the structural relaxation time under steady shear.

\begin{figure*}
    \centering
    \includegraphics[angle=0, width=\textwidth]{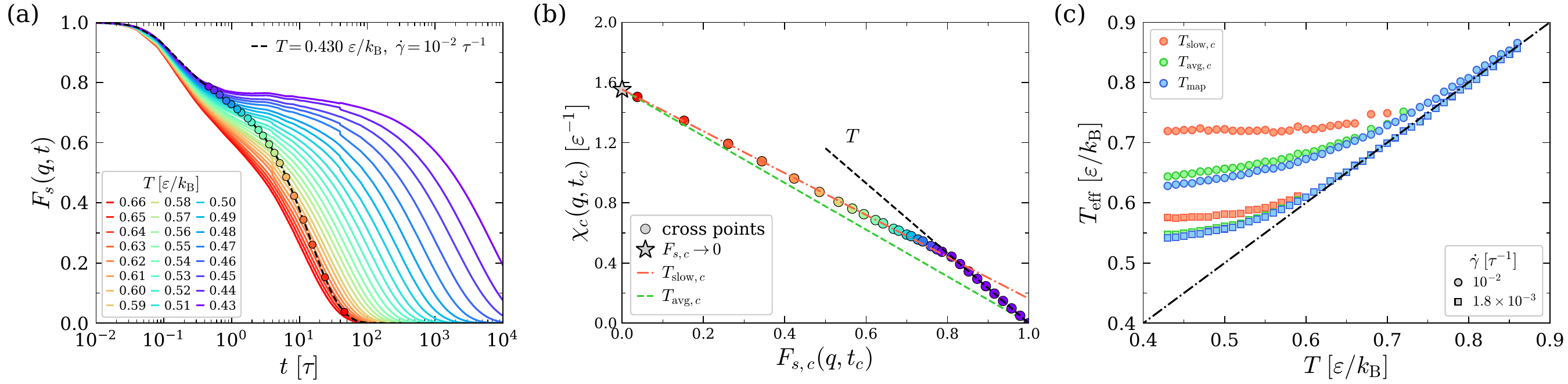}
    \caption{Determination of the effective temperature from the fluctuation-dissipation relation. (a) Self-intermediate scattering function $F_s(q,t,T,\dot{\gamma})$ at $T=0.43\,\varepsilon/k_{\mathrm{B}}$ and $\dot{\gamma}=10^{-2}\,\tau^{-1}$, compared with equilibrium results at different $T$. Circles mark the intersections between the nonequilibrium and equilibrium curves used to define the crossing points at $t\equiv t_c$. (b) Parametric plot $\chi_c(q,t_c)$ versus $F_{s,c}(q,t_c)$ constructed from the intersection points shown in (a). The slope of the slow-relaxation branch defines $T_{\mathrm{slow},c}$, while $T_{\mathrm{avg},c}$ is determined from the long-time endpoint $F_{s,c}\rightarrow0$; equivalently, $-1/(k_{\mathrm B}T_{\mathrm{avg},c})$ is the slope of the secant connecting $(F_{s,c},\chi_c)=(1,0)$ to this endpoint.
    (c) Effective temperatures $T_{\mathrm{map}}$, $T_{\mathrm{slow},c}$, and $T_{\mathrm{avg},c}$ as a function of the bath temperature $T$. The dash-dotted line indicates the equilibrium relation $T_{\mathrm{eff}}=T$. Similar behavior is observed for the polymer systems (see the SM).}
    \label{Fig4}
\end{figure*}

To continue, we focus on the density–density response, where $C(t)$ is identified with the intermediate scattering function $F_s(q,t)$. Figure~\ref{Fig4}(a) compares $F_s(q,t,T,\dot{\gamma})$ for a sheared fluid at $T=0.43 \,\varepsilon/k_{\mathrm{B}}$ and $\dot{\gamma}=10^{-2} \,\tau^{-1}$ with equilibrium data at different temperatures. At short times, the decay of $F_s(q,t)$ under shear closely follows the equilibrium curve at the bath temperature $T$, indicating that fast, local dynamics remain effectively thermalized and only weakly affected by the imposed flow.

At longer times, however, $F_s(q,t,T,\dot{\gamma})$ progressively deviates from the equilibrium bath-temperature curve. Crucially, this deviation cannot be captured by any single equilibrium curve at a higher temperature. Instead, the nonequilibrium correlator sequentially intersects a family of equilibrium curves corresponding to increasing temperatures, generating crossing points defined by
\begin{equation}
F_{s,c}(q,t_c,T_c,0)=F_s(q,t_c,T,\dot{\gamma}).
\label{cross}
\end{equation}
Here, the left-hand side denotes the equilibrium intermediate scattering function evaluated at temperature $T_c$, while the right-hand side corresponds to the driven system. This construction provides a dynamical mapping between the nonequilibrium system and a family of equilibrium states.

In the asymptotic limit $F_s(q,t,T,\dot{\gamma})\to 0$, the driven dynamics converge to a unique equilibrium curve. This highlights that any effective temperature extracted from the dynamics is inherently scale-dependent, interpolating from the bath temperature at short times to a larger asymptotic value associated with structural relaxation.

After identifying the crossing points in Eq.~\eqref{cross} [Fig.~\ref{Fig4}(a)], we define the corresponding equilibrium susceptibility as
\begin{equation}
\chi_c(q,t_c)\equiv \frac{1-F_{s,c}(q,t_c)}{k_{\mathrm B} T_c},
\end{equation}
where $F_{s,c}(q,t_c)$ is evaluated at the crossing condition.

We then apply the fluctuation-dissipation relation of Eq.~\eqref{ee2} to the pair $(\chi_c, F_{s,c})$, using $t_c$ as the parameter. A linear relation between $\chi_c(q,t_c)$ and $F_{s,c}(q,t_c)$ identifies an effective temperature through its slope. Unlike conventional analyses, the temperature is extracted from the crossing points rather than directly from the nonequilibrium correlation functions.

Figure~\ref{Fig4}(b) shows the corresponding parametric plot for the KABLJ system. As in previous nonequilibrium fluctuation-dissipation analyses, two distinct linear regimes are observed. For $F_{s,c}(q,t_c)\gtrsim 0.8$, the slope yields the bath temperature $T$, indicating that fast vibrational modes remain effectively equilibrated and only weakly affected by shear, consistent with the short-time behavior of $F_s(q,t,T,\dot{\gamma})$. For $F_{s,c}(q,t_c)\lesssim 0.8$, a second linear regime emerges, defining an effective temperature associated with the slow structural dynamics, $T_{\mathrm{slow},c}>T$. This temperature increases monotonically with shear rate, reflecting the progressively stronger departure from equilibrium induced by the external driving.

However, $T_{\mathrm{slow},c}$ does not exactly coincide with the phenomenological mapping temperature $T_{\mathrm{map}}$ introduced in the main text [Fig.~\ref{Fig4}(c)], and it systematically deviates from $T_{\mathrm{map}}$ as the shear rate $\dot{\gamma}$ increases. As a result, this definition fails to reproduce the shear-dependent structural relaxation time. This discrepancy is not specific to the KABLJ system, but persists across all three glass-forming systems considered in this work (see the SM).

We therefore turn to the alternative definition in Eq.~\eqref{eeav}, based on the asymptotic structure of the fluctuation–dissipation relation. In the long-time limit, where fast relaxation has fully decayed and $F_{s,c}(q,t_c)\rightarrow 0$, we define
\begin{equation}
k_{\mathrm B}T_{\mathrm{avg},c} \equiv \frac{1}{\chi_c(t_c \rightarrow \infty)},
\label{Eq_TFavg}
\end{equation}
which is determined by the long-time endpoint of the \(\chi_c\)--\(F_{s,c}\) curve. Equivalently, \(-1/(k_{\mathrm B}T_{\mathrm{avg},c})\) is the slope of the secant connecting the initial point \((F_{s,c},\chi_c)=(1,0)\) to the endpoint \(F_{s,c}\rightarrow0\), as illustrated by the dashed green line in Fig.~\ref{Fig4}(b).

As shown in Fig.~\ref{Fig4}(c) for the KABLJ system, $T_{\mathrm{avg},c}$ is in agreement with the phenomenological temperature $T_{\mathrm{map}}$ introduced in the main text. This establishes a direct link between $T_{\mathrm{map}}$ and a generalized fluctuation–dissipation construction, rather than a purely phenomenological mapping procedure. A similarly good agreement is observed in polymer glass-formers (see the SM), indicating that this correspondence is robust across a broad class of driven amorphous materials.

\end{document}


\title{Supplemental Material for ``String Model Predicts Relaxation in Sheared Glass-Forming Liquids''}

\author{Zi-Long Wang}
\affiliation{State Key Laboratory of Polymer Science and Technology, Changchun Institute of Applied Chemistry, Chinese Academy of Sciences, Changchun 130022, P. R. China}
\affiliation{School of Applied Chemistry and Engineering, University of Science and Technology of China, Hefei 230026, P. R. China}

\author{Qi-Lu Yuan}
\affiliation{State Key Laboratory of Polymer Science and Technology, Changchun Institute of Applied Chemistry, Chinese Academy of Sciences, Changchun 130022, P. R. China}
\affiliation{School of Applied Chemistry and Engineering, University of Science and Technology of China, Hefei 230026, P. R. China}

\author{Yun-Jiang Wang}
\affiliation{State Key Laboratory of Nonlinear Mechanics, Institute of Mechanics, Chinese Academy of Sciences, Beijing 100190, P. R. China}
\affiliation{School of Engineering Science, University of Chinese Academy of Sciences, Beijing 100049, P. R. China}

\author{Jack F. Douglas}
\email{jack.douglas@nist.gov}
\affiliation{Materials Science and Engineering Division, National Institute of Standards and Technology, Gaithersburg, Maryland 20899, United States}

\author{Matteo Baggioli}
\email{b.matteo@sjtu.edu.cn}
\affiliation{Wilczek Quantum Center, School of Physics and Astronomy, Shanghai Jiao Tong University, Shanghai 200240, P. R. China}
\affiliation{Shanghai Research Center for Quantum Sciences, Shanghai 201315, P. R. China}

\author{Zhao-Yan Sun}
\email{zysun@ciac.ac.cn}
\affiliation{State Key Laboratory of Polymer Science and Technology, Changchun Institute of Applied Chemistry, Chinese Academy of Sciences, Changchun 130022, P. R. China}
\affiliation{School of Applied Chemistry and Engineering, University of Science and Technology of China, Hefei 230026, P. R. China}
\affiliation{Jilin Provincial International Cooperation Key Laboratory for Polymer Processing Physics, Changchun 130022, P. R. China}

\author{Wen-Sheng Xu}
\email{wsxu@nju.edu.cn}
\affiliation{MOE Key Laboratory of High Performance Polymer Materials and Technology, State Key Laboratory of Coordination Chemistry, School of Chemistry, Nanjing University, Nanjing 210023, P. R. China}

\date{\today}

\maketitle

\tableofcontents

\section{Model Systems and Simulation Details}

\subsection{Kob-Andersen Glass-Forming Liquid}
\label{Sec_KA}

We consider a canonical model of glass-forming liquids, namely the Kob--Andersen (KA) binary Lennard--Jones (LJ) liquid \cite{Scaling_1994_73_1376, Testing_1995_51_4626}, which we denote as KABLJ throughout this work. The KA model consists of two particle species, $A$ and $B$, of equal mass $m$, with a composition ratio $N_A:N_B=4\mathbin{:}1$. Particles interact via a truncated-and-shifted LJ potential,
\begin{eqnarray}
	\label{Eq_LJ}
	U_{\mathrm{LJ}, \alpha\beta}(r) = 4\varepsilon_{\alpha\beta} \left[ \left(\frac{\sigma_{\alpha\beta}}{r}\right)^{12} - \left(\frac{\sigma_{\alpha\beta}}{r}\right)^6 \right] + C_{\alpha\beta}(r_{\mathrm{cut}}),
\end{eqnarray}
for $r<r_{\mathrm{cut}}$, and zero otherwise. Here, $\alpha,\beta\in\{A,B\}$, $r$ denotes the distance between two particles, and $\sigma_{\alpha\beta}$ and $\varepsilon_{\alpha\beta}$ set the characteristic length and energy scales, respectively. The constant $C_{\alpha\beta}(r_{\mathrm{cut}})$ ensures that $U_{\mathrm{LJ},\alpha\beta}(r)$ vanishes continuously at the cutoff distance $r_{\mathrm{cut}}=2.5\,\sigma_{\alpha\beta}$. The interaction parameters are $\varepsilon_{AB}/\varepsilon_{AA}=1.5$, $\varepsilon_{BB}/\varepsilon_{AA}=0.5$, $\sigma_{AB}/\sigma_{AA}=0.8$, and $\sigma_{BB}/\sigma_{AA}=0.88$.

Results are reported in standard reduced LJ units. Specifically, the units of length, time, and temperature are $\sigma=\sigma_{AA}$, $\tau=\sqrt{m\sigma_{AA}^2/\varepsilon_{AA}}$, and $\varepsilon/k_{\mathrm B}=\varepsilon_{AA}/k_{\mathrm B}$, respectively, where $k_{\mathrm B}$ is the Boltzmann constant. We introduce the symbols $\sigma$ and $\varepsilon$ as shorthand for the length and energy scales of the $A$ particles to maintain a common notation for both the KABLJ and polymer systems.

The total number of particles is $N=N_A+N_B=4000$. Following the standard protocol for the KABLJ model \cite{Scaling_1994_73_1376, Testing_1995_51_4626}, we investigate glass formation under constant-volume conditions at a number density $\rho=N/V=1.2 \ \sigma^{-3}$.

\subsection{Coarse-Grained Polymer Melt}
\label{Sec_Polymer}

We also consider a coarse-grained bead-spring model of polymer melts \cite{Dynamics_1990_92_5057, Molecular_1986_33_3628}. Each polymer chain consists of a sequence of connected beads. Bond connectivity between neighboring beads is maintained by the finitely extensible nonlinear elastic (FENE) potential \cite{Dynamics_1990_92_5057, Molecular_1986_33_3628},
\begin{eqnarray}
	\label{Eq_FENE}
	U_{\mathrm{FENE}}(r) = -\frac{1}{2} k_b R_0^2 \ln\left[1-\left(\frac{r}{R_0}\right)^2\right] + 4 \varepsilon \left[ \left(\frac{\sigma}{r}\right)^{12} - \left(\frac{\sigma}{r}\right)^6 \right] + \varepsilon,
\end{eqnarray}
where $\varepsilon$ and $\sigma$ are the characteristic energy and length scales of the LJ interaction. The first term in Eq.~\ref{Eq_FENE} is defined up to $R_0$, while the second term is truncated at $2^{1/6} \ \sigma$. We adopt the standard parameter values $k_b = 30 \ \varepsilon/\sigma^2$ and $R_0 = 1.5 \ \sigma$. Non-bonded interactions are described by Eq.~\ref{Eq_LJ}, where the subscript $\alpha\beta$ is omitted for the one-component melt and the cutoff distance is $r_{\mathrm{cut}} = 2.5 \ \sigma$.

As for the KABLJ system, all quantities and results for the polymer melt are expressed in standard reduced LJ units. Specifically, length, time, temperature, and pressure are measured in units of $\sigma$, $\tau$, $\varepsilon/k_{\mathrm B}$, and $\varepsilon/\sigma^3$, respectively, where $\tau=\sqrt{m\sigma^2/\varepsilon}$. Each chain contains $20$ beads of equal mass $m$, and the melt consists of $400$ chains, yielding a total bead number of $N=8000$.

For the polymer melt system, we study glass formation under both constant-volume and constant-pressure conditions. The constant-volume system is simulated at a number density $\rho=1.0 \ \sigma^{-3}$, whereas the constant-pressure system is simulated at $P=0.0\,\varepsilon/\sigma^3$. For convenience, we refer to these systems as Poly-V and Poly-P, respectively.

\subsection{Simulation Details}
\label{Sec_Simulation}

Our simulations are carried out in three dimensions using the LAMMPS simulation package \cite{Fast_1995_117_1, LAMMPS_webpage}. The equations of motion are integrated with a time step of $\Delta t = 0.005\,\tau$. We follow the protocol established in our previous studies \cite{Molecular_2020_53_4796, Role_2020_53_9678, Investigation_2020_53_6828a} to investigate glass formation in the absence of shear. The system is first fully equilibrated at $T=2.0 \ \varepsilon/k_{\mathrm B}$ for the KABLJ and Poly-V systems at the specified density, or for the Poly-P system at $P=0.0\,\varepsilon/\sigma^3$. The equilibrated liquid is then cooled at a rate of $10^{-4}\,\varepsilon/(k_{\mathrm B}\tau)$ under periodic boundary conditions in either the $NVT$ or $NPT$ ensemble, depending on the thermodynamic conditions considered. Temperature and pressure are controlled using the Nos{\'e}--Hoover thermostat and barostat implemented in LAMMPS.

To obtain properties at a given temperature, configurations generated during the cooling process are further equilibrated for a duration typically $10$--$100$ times longer than the segmental structural relaxation time in the $NVT$ ensemble, irrespective of the thermodynamic path used to generate the state point. We focus on temperatures well above the glass transition temperature $T_{\mathrm g}$, such that nonequilibrium effects associated with the glassy state are irrelevant to the present study.

After equilibration, shear deformation is applied at a constant shear rate $\dot{\gamma}$ following standard procedures \cite{Predictive_2020_6_eaaz0777, Probing_2017_114_7952, Rheological_2019_67_66, Structural_2018_115_87}. Specifically, simple shear is imposed in the $xy$ plane under constant-volume conditions using the SLLOD equations of motion \cite{Nonlinear_1984_30_1528, Simple_2006_124_194103} in conjunction with Lees--Edwards boundary conditions. In this geometry, the $x$, $y$, and $z$ directions correspond to the flow, velocity-gradient, and vorticity directions, respectively. We vary $\dot{\gamma}$ from $10^{-5} \ \tau^{-1}$ to $10^{-2} \ \tau^{-1}$ over a wide range of temperatures. To ensure that the steady state is reached, the system is deformed up to a total strain $\gamma = \dot{\gamma} t \approx 100$ for all state points studied, which is sufficient according to previous simulations of the same polymer model \cite{Predictive_2020_6_eaaz0777}. Static and dynamic properties are subsequently computed from trajectories collected in the steady-shear regime.

\begin{figure}[htbp]
	\centering
	\includegraphics[angle=0, width=0.6\textwidth]{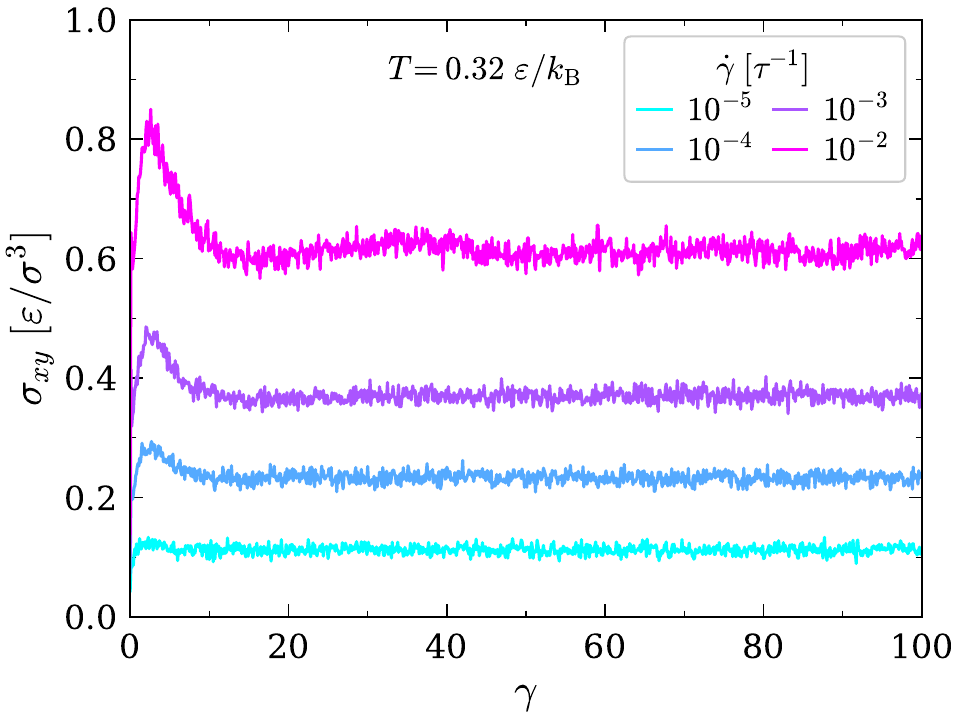}
	\caption{Shear stress under steady shear in the Poly-V system. Shear stress $\sigma_{xy}$ as a function of shear strain $\gamma$ for different shear rates $\dot{\gamma}$ at $T=0.32 \ \varepsilon/k_{\mathrm B}$. Similar results are obtained for the KABLJ and Poly-P systems.}
	\label{S1}
\end{figure}

As an illustration, Fig.~\ref{S1} shows the shear stress $\sigma_{xy}$ as a function of shear strain $\gamma$ for different shear rates $\dot{\gamma}$ at $T=0.32 \ \varepsilon/k_{\mathrm B}$ in the Poly-V system. The stress initially increases with increasing strain and exhibits a characteristic overshoot before decreasing toward a steady-state value. At larger strains, the system reaches a stationary flowing state in which time-independent average properties are established. Evidently, the maximum strain of $\gamma \approx 100$ is sufficiently large to ensure that all simulations are performed in the steady-shear regime.

\section{Definition of Basic Properties}
\label{Sec_Basic}

Here we describe how structural relaxation and stringlike cooperative motion, the two central quantities considered in this study, are quantified. We determine the structural relaxation time $\tau_{\alpha}$ from the self-part of the intermediate scattering function,
\begin{eqnarray}
	\label{Eq_FSQ}
	F_s(q,t) = \frac{1}{N} \left\langle \sum_{j=1}^{N} \exp\left( -i\mathbf{q}\cdot\left[\mathbf{r}_j(t)-\mathbf{r}_j(0)\right] \right) \right\rangle,
\end{eqnarray}
where $i$ is the imaginary unit, $q=|\mathbf{q}|$ is the wave number, $\mathbf{r}_j$ is the position of particle $j$, and $\langle \cdots \rangle$ denotes the usual thermal average. For the KABLJ system, $F_s(q,t)$ is computed using only the $A$ particles. The wave number is chosen as $q = 7.2 \ \sigma^{-1}$ for the KABLJ system and $q=7.0 \ \sigma^{-1}$ for the polymer systems, corresponding approximately to the first peak of the static structure factor $S(q)$. The structural relaxation time $\tau_{\alpha}$ is then defined as the time at which $F_s(q,t)$ decays to $0.2$.

To quantify stringlike cooperative motion, we follow the established procedure of Refs.~\citenum{Polymer_2003_119_5290, Relationship_2013_138_12A541}. Mobile particles are defined as a fraction $f_0$ of particles exhibiting the largest displacement over a chosen time interval, with $f_0=5.0\%$ and $6.5\%$ for the KABLJ and polymer systems, respectively. Two mobile particles $j$ and $k$ are considered to belong to the same string if the following condition holds,
\begin{eqnarray}
	\label{Eq_Min}
	\min \left[|\mathbf{r}_j(t)-\mathbf{r}_k(0)|, |\mathbf{r}_k(t)-\mathbf{r}_j(0)|\right] < \delta,
\end{eqnarray}
where $\delta$ is a displacement threshold set to $0.6 \ \sigma$ and $0.55 \ \sigma$ for the KABLJ and polymer systems, respectively. The number-averaged string length is computed as $\langle s(t)\rangle=\langle \sum_{s=1}^{\infty}sC(s)\rangle/\sum_{s=1}^{\infty}C(s)$, where $C(s)$ is the probability of observing a string of length $s$. The quantity $\langle s(t)\rangle$ exhibits a maximum $L\equiv\langle s(t_L)\rangle$ at a characteristic time $t_L$, which defines the characteristic string length $L$. Note that all particles are included in the calculation of $L$, including both particle species in the KABLJ system.

In the presence of shear flow, the calculation of $\tau_{\alpha}$ and $L$ is complicated by the need to subtract the affine deformation induced by the flow. Reference~\cite{Dynamics_1998_58_3515} showed that the displacement vector appearing in Eq.~\ref{Eq_FSQ} should be replaced by
\begin{eqnarray}
	\label{Eq_Distacne_Shear}
	\Delta \mathbf{r}_j(t) = \mathbf{r}_j(t)-\mathbf{r}_j(0) - \dot{\gamma} \int_{0}^{t} dt'\, y_j(t')\mathbf{x},
\end{eqnarray}
where $y_j(t')$ is the $y$ component of the position of particle $j$ and $\mathbf{x}$ is a unit vector in the flow direction. Computing the full integral in Eq.~\ref{Eq_Distacne_Shear} requires extensive bookkeeping of particle trajectories and is therefore computationally demanding. Fortunately, Tanaka and co-workers \cite{Anisotropic_2009_102_016001, Structural_2018_115_87} demonstrated that
\begin{eqnarray}
	\label{Eq_Distacne_Approx}
	\dot{\gamma}\int_{0}^{t}dt'\,y_j(t') \approx \dot{\gamma}t\,y_j(0),
\end{eqnarray}
provides an excellent approximation, which we adopt in the present work to compute both $F_s(q,t)$ and $\langle s(t)\rangle$.

\begin{figure}[htbp]
	\centering
	\includegraphics[angle=0, width=0.975\textwidth]{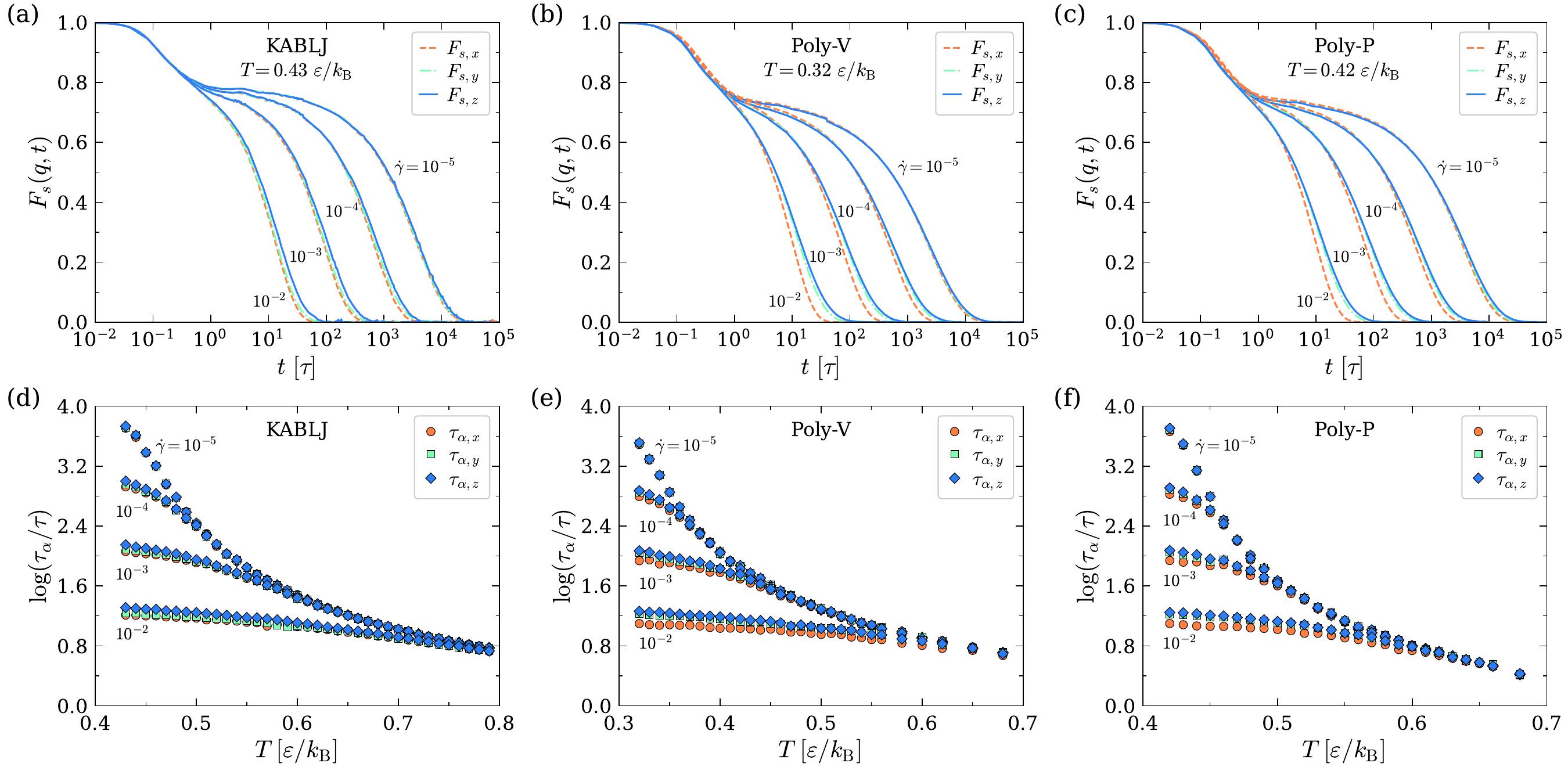}
	\caption{Dynamics in different directions under steady shear. (a--c) $F_s(q,t)$ in the $x$, $y$, and $z$ directions for the three systems at different $\dot{\gamma}$ and a fixed $T$ indicated. (d--f) Logarithm of the structural relaxation time, $\log(\tau_\alpha)$, in the $x$, $y$, and $z$ directions as a function of $T$ for the KABLJ system for the three systems at different $\dot{\gamma}$.}
	\label{Fig_Shear_FSQ}
\end{figure}

Previous studies have also shown that $F_s(q,t)$ exhibits only weak anisotropy under steady shear \cite{Dynamics_1998_58_3515, Anisotropic_2009_102_016001}. We confirm this finding in our simulations (Figure~\ref{Fig_Shear_FSQ}). The small differences in the $x$ direction at high shear rates are expected since the approximation of Eq.~\ref{Eq_Distacne_Approx} is less accurate under such conditions. For this reason, we consider only $F_{s,z}(q,t)$ along the shear-free ($z$) direction \cite{Predictive_2020_6_eaaz0777}, i.e., $F_s(q,t)=F_{s,z}(q,t)$.

\section{Results at Equilibrium}
\label{Sec_Equil}

\begin{figure}[htbp!]
	\centering
	\includegraphics[angle=0, width=0.95\textwidth]{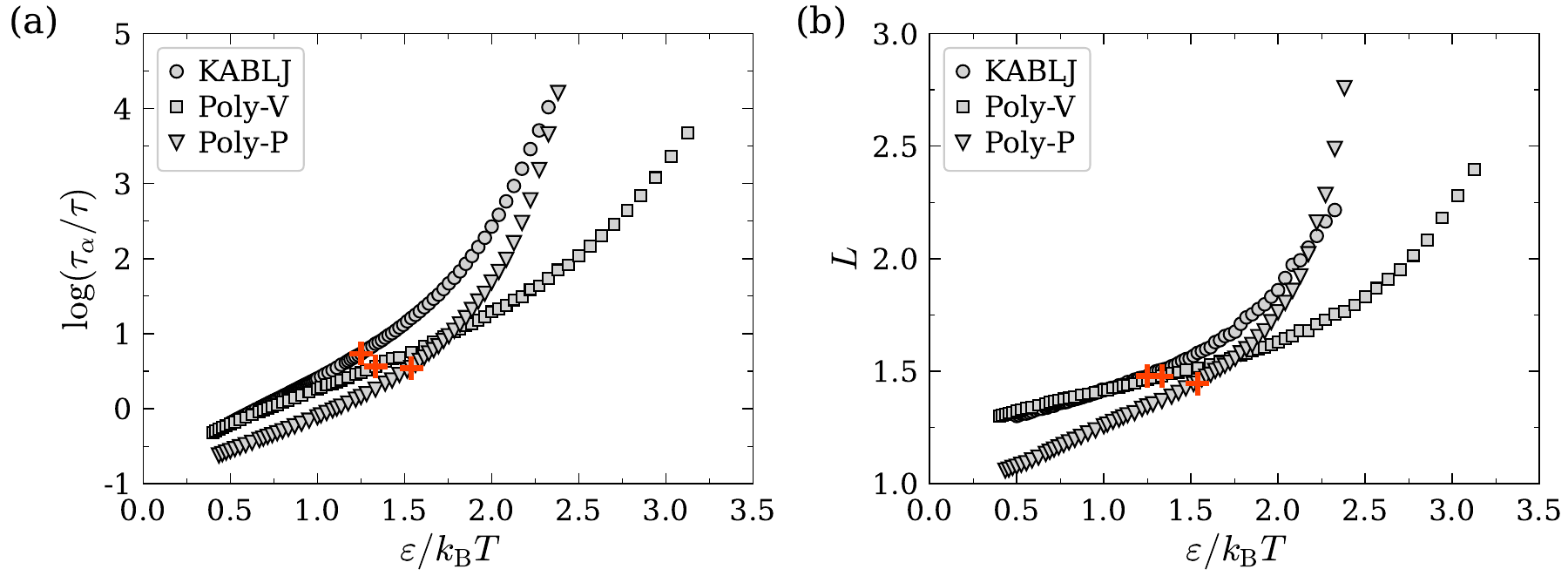}
	\caption{Temperature dependence of the characteristic properties of glass formation. (a,b) Logarithm of the structural relaxation time $\log(\tau_{\alpha})$ and average string length $L$ versus inverse temperature $\varepsilon / k_{\mathrm{B}} T$ for the three systems, respectively. The plus symbols indicate the positions of the onset temperature $T_A$ of glass formation.}
\label{S2}
\end{figure}

This section characterizes the equilibrium glass-forming behavior of the three systems studied in this work. We begin by examining the temperature dependence of the structural relaxation time $\tau_{\alpha}$ and the characteristic string length $L$. As shown in Fig.~\ref{S2}, structural relaxation slows dramatically upon cooling, accompanied by a growth in the extent of stringlike cooperative motion. Such behavior is characteristic of glass-forming liquids.

The characteristic temperatures associated with glass formation are determined using established procedures. The onset temperature $T_A$ of non-Arrhenius dynamics is estimated from the appearance of a minimum near $t=2 \, \tau$ in the logarithmic derivative of the mean-squared displacement \cite{Relationship_2013_138_12A541}. Because the timescale defining $T_A$ is closely related to particle localization, $T_A$ is often referred to as the localization temperature. The resulting values of $T_A$ are indicated by the plus symbols in Fig.~\ref{S2}(a) and are listed in Table~I of the main text.

\begin{figure}[htbp!]
	\centering
	\includegraphics[angle=0, width=0.95\textwidth]{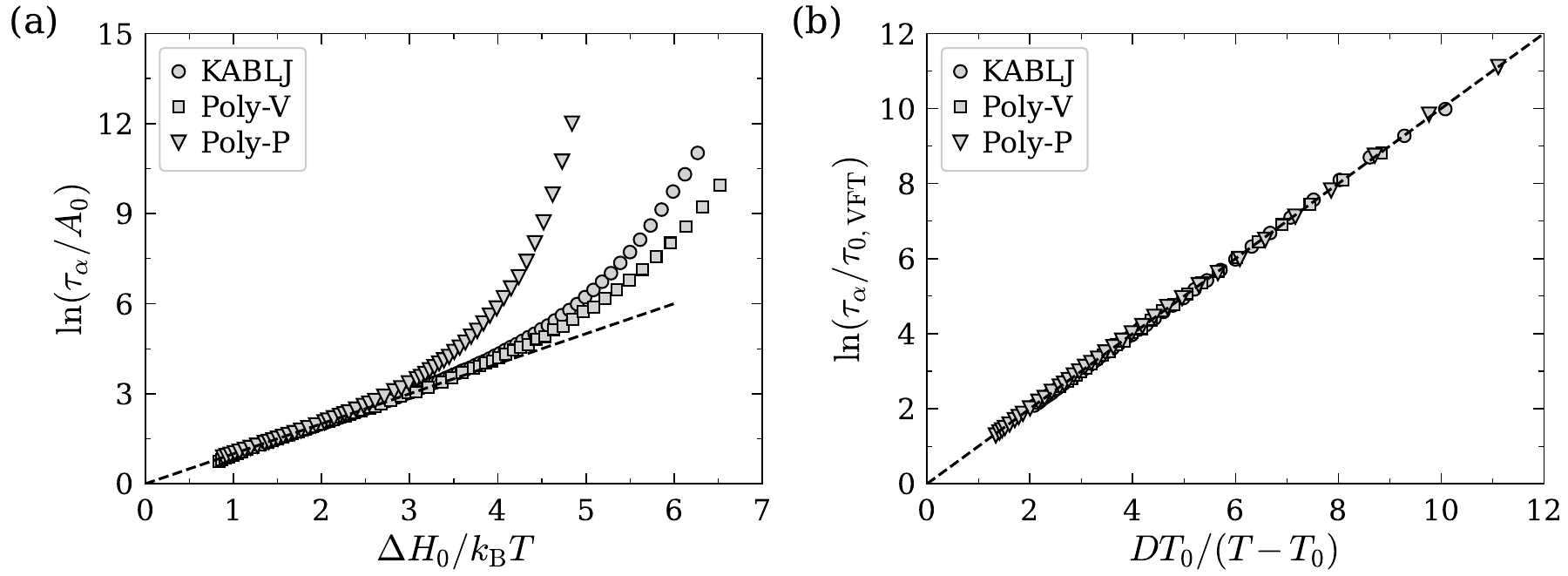}
	\caption{Analysis of the temperature dependence of the structural relaxation time. (a,b) Arrhenius dynamics (Eq. \ref{Eq_Arrhenius}) in the high-$T$ regime and Vogel--Fulcher--Tammann (VFT) equation (Eq. \ref{Eq_VFT}) in the low-$T$ regime for the three systems, respectively, where dashed lines indicate $\ln(\tau_{\alpha}/A_{0}) = \Delta H_0 / k_{\mathrm{B}} T$ and $\ln(\tau_{\alpha} / \tau_{0, \mathrm{VFT}}) = D T_0 / (T - T_0)$. The meanings of $A_{0}$, $ \Delta H_0$, $\tau_{0, \mathrm{VFT}}$, $D$, and $T_0$ are explained in the text.}
\label{S3}
\end{figure}

An alternative estimate of the onset of non-Arrhenius relaxation can be obtained directly from the temperature dependence of $\tau_{\alpha}$. Figure~\ref{S3}(a) shows that our simulations cover a temperature range in which the high-temperature dynamics follow the Arrhenius form expected from transition state theory (TST) \cite{Theory_1941_28_301, Book_Eyring},
\begin{eqnarray}
	\label{Eq_Arrhenius}
	\tau_{\alpha} = A_0 \exp \left(\frac{\Delta H_0} {k_{\mathrm{B}} T} \right),
\end{eqnarray}
where $A_0$ is a prefactor and $\Delta H_0$ is the activation energy. The prefactor $A_0$ contains both a vibrational attempt frequency and an entropic contribution proportional to $\exp(-\Delta S_0/k_{\mathrm B})$, where $\Delta S_0$ is the activation entropy.

\begin{table}[htbp!]
	\centering
	\caption{Characteristic properties of glass formation for the three systems at equilibrium. $T_c$ and $T_0$ are the crossover and ideal glass transition temperatures, respectively. $K_{\mathrm{VFT}}$ is the VFT fragility parameter.}
	\label{Table_Basic}
	\begin{ruledtabular}
	\begin{tabular}{lccc}
	System & $T_c \ [\varepsilon / k_{\mathrm{B}}]$ & $T_0 \ [\varepsilon / k_{\mathrm{B}}]$ & $K_{\mathrm{VFT}}$ \\
	\colrule
	KABLJ & $0.424$ & $0.313$ & $0.264$ \\
	Poly-V & $0.298$ & $0.213$ & $0.224$ \\
	Poly-P & $0.413$ & $0.347$ & $0.430$ \\
	\end{tabular}
	\end{ruledtabular}
\end{table}

The crossover temperature $T_c$ is obtained by fitting simulation data in the range $5\tau < \tau_{\alpha} < 5\times10^3 \, \tau$ to the power law $\tau_{\alpha}\sim(T-T_c)^{-\gamma}$, where $\gamma$ is a fitting parameter. This scaling form was originally proposed within mode-coupling theory (MCT) \cite{Book_Gotze}, and the fitted parameter $T_c$ is therefore commonly referred to as the mode-coupling temperature. The resulting values of $T_c$ are listed in Table~\ref{Table_Basic}.

\begin{figure}[htbp!]
	\centering
	\includegraphics[angle=0, width=0.6\textwidth]{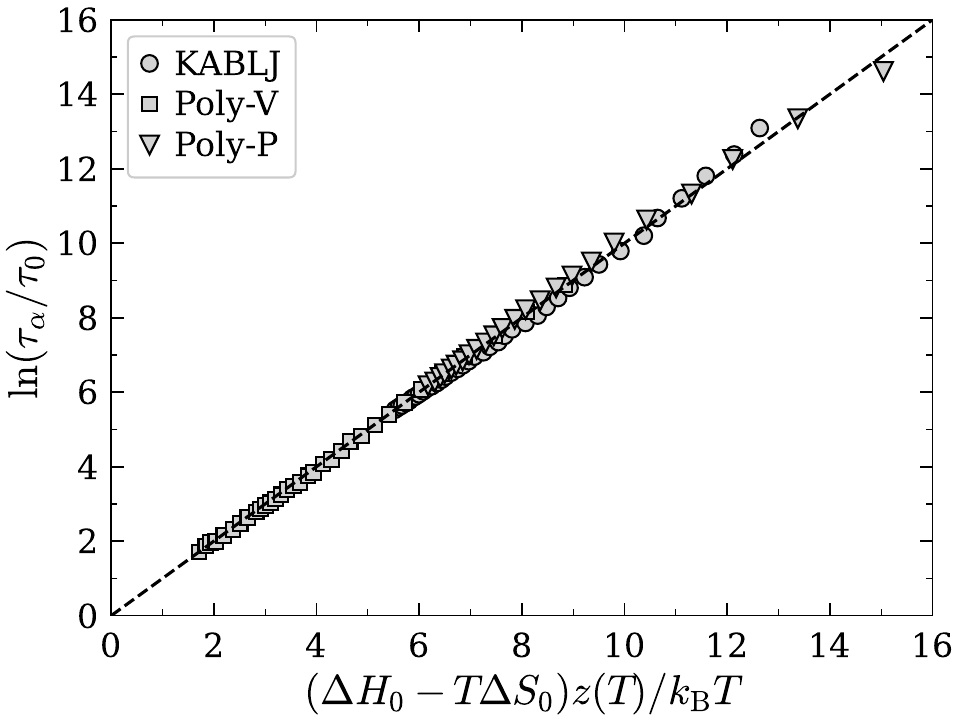}
	\caption{String Model description of glass formation at equilibrium. Logarithm of the structural relaxation time, $\ln(\tau_{\alpha}/\tau_{0})$, as a function of $(\Delta H_{0}-T\Delta S_{0})z(T) / k_{\mathrm B}T$ for the three glass-forming systems. The parameters $\tau_{0}$, $\Delta H_{0}$, and $\Delta S_{0}$ are reported in Table~1 of the main text. The dashed line indicates the predicted one-to-one correspondence between the two quantities.}
	\label{S12}
\end{figure}

The ideal glass transition temperature $T_0$ is estimated using the Vogel--Fulcher--Tammann (VFT) equation \cite{Law_1921_22_645, Analysis_1925_8_339a, Die_1926_156_245},
\begin{eqnarray}
	\label{Eq_VFT}
	\tau_{\alpha} = \tau_{0, \mathrm{VFT}}\exp\left(\frac{D T_0}{T - T_0}\right),
\end{eqnarray}
where $\tau_{0,\mathrm{VFT}}$ is a prefactor and $D$ is the fragility parameter that characterizes the temperature dependence of $\tau_{\alpha}$. As shown in Fig.~\ref{S3}(b), the VFT expression provides an excellent description of the non-Arrhenius regime. We further define $K_{\mathrm{VFT}}\equiv1/D$, such that larger values of $K_{\mathrm{VFT}}$ correspond to more fragile glass-forming liquids. The fitted values of both $T_0$ and $K_{\mathrm{VFT}}$ are reported in Table~\ref{Table_Basic}.

Finally, we show in Fig.~\ref{S12} that the String Model provides an excellent description of our simulation data at equilibrium for all the three systems. We note that $\Delta S_0$ is a kinetic parameter describing barrier crossing events in an energy landscape so that it is not related in any obvious way to the equilibrium fluid entropy. Hence, it is understandable both from a theoretical standpoint and from measurements since the earliest formulations of transition state theory that $\Delta S_0$ can be either positive or negative. We also note that the String Model works well even when the entropic contribution to the activation free energy is neglected, as shown in a recent work~\cite{Yuan_2026_13_nwag366}.

\section{Results under Steady Shear}
\label{Sec_Steady}

\subsection{\label{Sec_Steady_Characteristic}Characteristic Properties of Glass Formation}

\begin{figure}[htbp!]
	\centering
	\includegraphics[angle=0, width=0.9\textwidth]{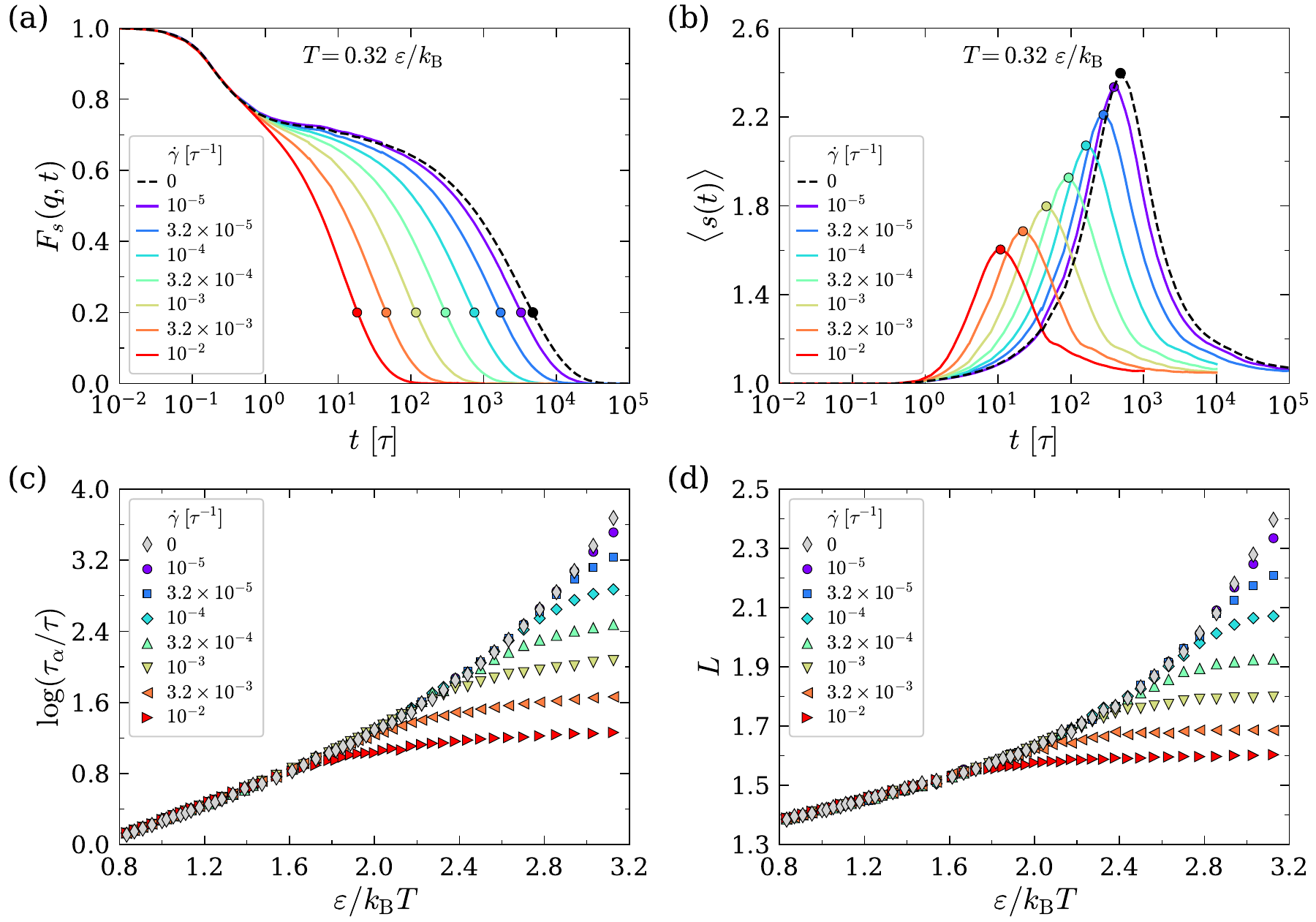}
	\caption{Structural relaxation and stringlike cooperative motion in the Poly-V system under steady shear. (a,b) Self-intermediate scattering function $F_s(q,t)$ and number-averaged string length $\langle s(t)\rangle$ as a function of time $t$ for different shear rates $\dot{\gamma}$ at $T=0.32\,\varepsilon/k_{\mathrm B}$, respectively. (c,d) Structural relaxation time $\log(\tau_\alpha)$ and characteristic string length $L$ as a function of inverse temperature $\varepsilon/k_{\mathrm B}T$, respectively.}
	\label{S10}
\end{figure}

\begin{figure}[htbp!]
	\centering
	\includegraphics[angle=0, width=0.9\textwidth]{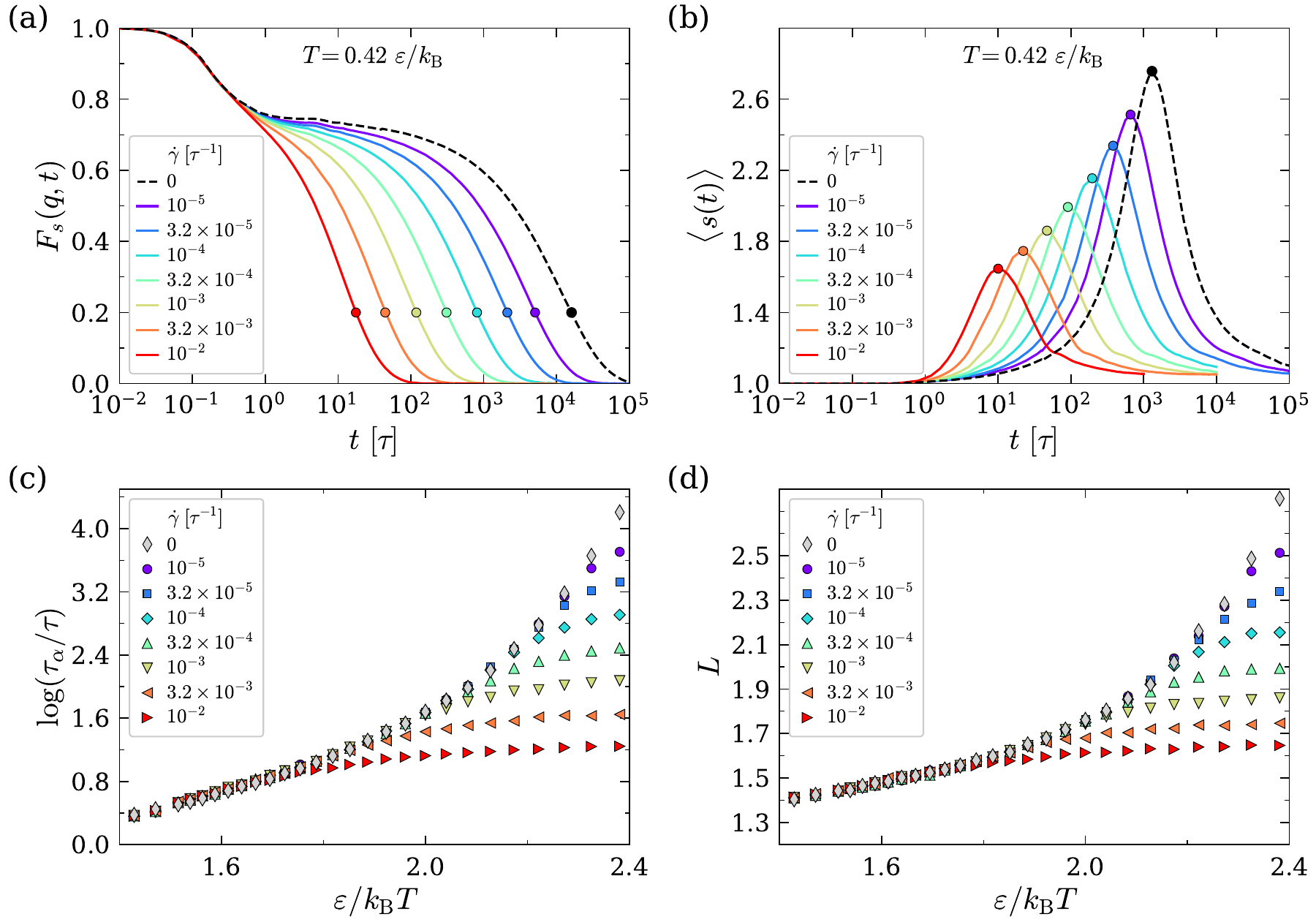}
	\caption{Structural relaxation and stringlike cooperative motion in the Poly-P system under steady shear. (a,b) Self-intermediate scattering function $F_s(q,t)$ and number-averaged string length $\langle s(t)\rangle$ as a function of time $t$ for different shear rates $\dot{\gamma}$ at $T=0.42\,\varepsilon/k_{\mathrm B}$, respectively. (c,d) Structural relaxation time $\log(\tau_\alpha)$ and characteristic string length $L$ as a function of inverse temperature $\varepsilon/k_{\mathrm B}T$, respectively.}
	\label{S11}
\end{figure}

\begin{figure}[htbp!]
	\centering
	\includegraphics[angle=0, width=0.9\textwidth]{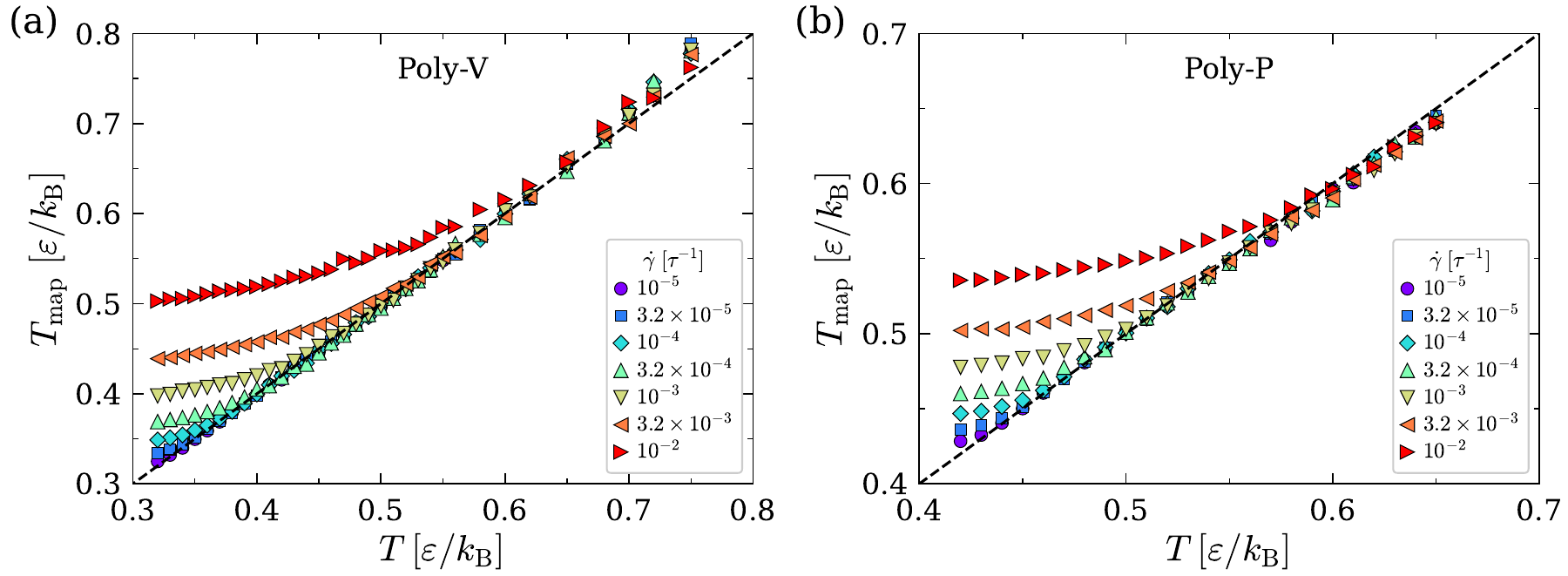}
	\caption{Summary of the effective temperature for the polymer systems. (a,b) Effective temperature $T_{\mathrm{map}}$ as a function of the bath temperature $T$ for the Poly-V and Poly-P systems over a wide range of shear rates $\dot{\gamma}$, respectively. The dashed line indicates $T_{\mathrm{map}} = T$.}
	\label{S13}
\end{figure}

\begin{figure}[htbp!]
	\centering
	\includegraphics[angle=0, width=0.6\textwidth]{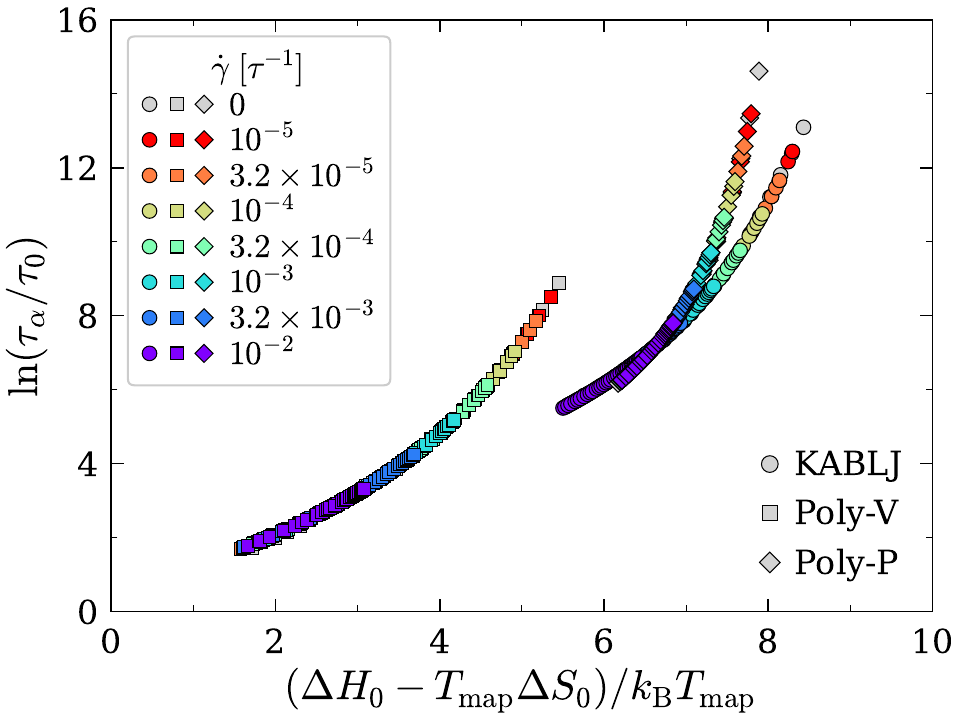}
	\caption{Dependence of the structural relaxation time on the effective temperature. The plot shows $\ln(\tau_{\alpha}/\tau_{0})$ as a function of $(\Delta H_{0}-T_{\mathrm{map}}\Delta S_{0}) / k_{\mathrm{B}}T_{\mathrm{map}}$ for all three systems and a wide range of shear rates.}
	\label{Fig_Shear_String1_Geff}
\end{figure}

In the main text, we present the results for structural relaxation and stringlike cooperative motion in the KABLJ system under steady shear. For completeness, the corresponding results for the Poly-V and Poly-P systems are shown in Figs.~\ref{S10} and~\ref{S11}, respectively. We find that all three systems exhibit qualitatively similar trends in both structural relaxation and stringlike cooperative motion as the shear rate is varied.

For completeness, we also summarize the effective temperature $T_{\mathrm{map}}$ for the polymer systems in Fig.~\ref{S13}, where we see that the results are similar to those for the KABLJ system.

To make our conclusion more robust, Fig.~\ref{Fig_Shear_String1_Geff} plots $\ln(\tau_{\alpha}/\tau_{0})$ as a function of $(\Delta H_{0}-T_{\mathrm{map}}\Delta S_{0}) / k_{\mathrm{B}}T_{\mathrm{map}}$. For each system, we see that the data for different $\dot{\gamma}$ collapse onto a single curve, which is expected since $T_{\mathrm{map}}$ is determined from the mapping of $\tau_{\alpha}$. Importantly, excluding the term $z(T)$ in the String Model does not lead to a data collapse across different systems, indicating that $T_{\mathrm{map}}$ alone is inadequate to predict the relaxation time. As stated in the main text, the effective temperature and the shear-induced suppression of the string length are the two ingredients that lead to a quantitative prediction of the structural relaxation time across a broad range of temperatures and shear rates. Our analysis thus eliminates the circularity issue that may arise from the determination of $T_{\mathrm{map}}$.

\newpage
\subsection{Effective Temperatures from Fluctuation-Dissipation Relation}
\label{Sec_Steady_FDR}

We follow the established methodology to determine an effective temperature associated with the slow modes of a sheared fluid. This approach is based on a nonequilibrium generalization of the fluctuation-dissipation theorem (FDT). We refer the reader to Ref.~\cite{Fluctuation_2000_63_012503} for a detailed discussion and summarize here only the essential steps required for its implementation.

We first define the following observables associated with density fluctuations,
\begin{eqnarray}
\label{Eq_A}
A(q,t)
=
\frac{1}{N}
\sum_j
\xi_j
\exp[i\,\mathbf{q}\cdot\mathbf{r}_j(t)],
\end{eqnarray}
and
\begin{eqnarray}
\label{Eq_B}
B(q,t)
=
2
\sum_j
\xi_j
\cos[\mathbf{q}\cdot\mathbf{r}_j(t)],
\end{eqnarray}
where $\xi_j=\pm1$ are random variables satisfying the constraint $\sum_j\xi_j=0$. The real part of the cross-correlation function of these observables is defined as
\begin{eqnarray}
\label{Eq_C}
C(q,t)
=
\langle A(q,t+t_0)B(q,t_0)\rangle
-
\langle A(q,t_0)\rangle
\langle B(q,t_0)\rangle,
\end{eqnarray}
which can be shown to be equivalent to the real part of the self-intermediate scattering function,
\begin{eqnarray}
\label{Eq_Fs}
C(q,t)
=
\frac{1}{N}
\left\langle
\sum_j
\cos[\mathbf{q}\cdot(\mathbf{r}_j(t)-\mathbf{r}_j(0))]
\right\rangle.
\end{eqnarray}
To probe the linear response, we apply a weak external field $h$ conjugate to $B(q,t)$ at time $t_0$, such that a perturbation $\Delta H=-hB(q,t)$ is added to the system Hamiltonian. The corresponding response function is
\begin{eqnarray}
\label{Eq_R}
R(q,t)
=
\left.
\frac{\delta\langle A(q,t+t_0)\rangle}
{\delta h(t_0)}
\right|_{h=0},
\end{eqnarray}
and the integrated susceptibility is defined as
\begin{eqnarray}
\label{Eq_chi_integral}
\chi(q,t)
=
\int_{t_0}^{t+t_0}
dt''
\left.
\frac{\partial\langle A(t)\rangle_h}
{\partial h(t'')}
\right|_{h\to0}.
\end{eqnarray}
For a sufficiently small constant perturbation $h$ switched on at time $t_0$, $\chi(q,t)$ can be approximated by
\begin{eqnarray}
\label{Eq_chi_approx}
\chi(q,t)
\approx
\frac{
\langle A_h(q,t+t_0)-A_{\mathrm{ref}}(q,t+t_0)\rangle
}{h}.
\end{eqnarray}
An effective temperature $T_{\mathrm{slow}}$ can then be introduced through the fluctuation-dissipation relation (FDR),
\begin{eqnarray}
\label{Eq_FDT}
R(q,t)
=
-\frac{1}{k_{\mathrm B}T_{\mathrm{slow}}}
\frac{\partial C(q,t)}{\partial t},
\end{eqnarray}
or, equivalently, through its integrated form,
\begin{eqnarray}
\label{Eq_FDT_integral}
\chi(q,t)
=
\int_{C(q,t)}^{C(q,0)}
\frac{dx}
{k_{\mathrm B}T_{\mathrm{slow}}(x)}.
\end{eqnarray}
A constant $T_{\mathrm{slow}}$ therefore appears as a linear slope of $-1/(k_{\mathrm B}T_{\mathrm{slow}})$ in a parametric plot of $\chi(q,t)$ versus $C(q,t)$.

\begin{figure}[htbp!]
	\centering
	\includegraphics[angle=0, width=0.975\textwidth]{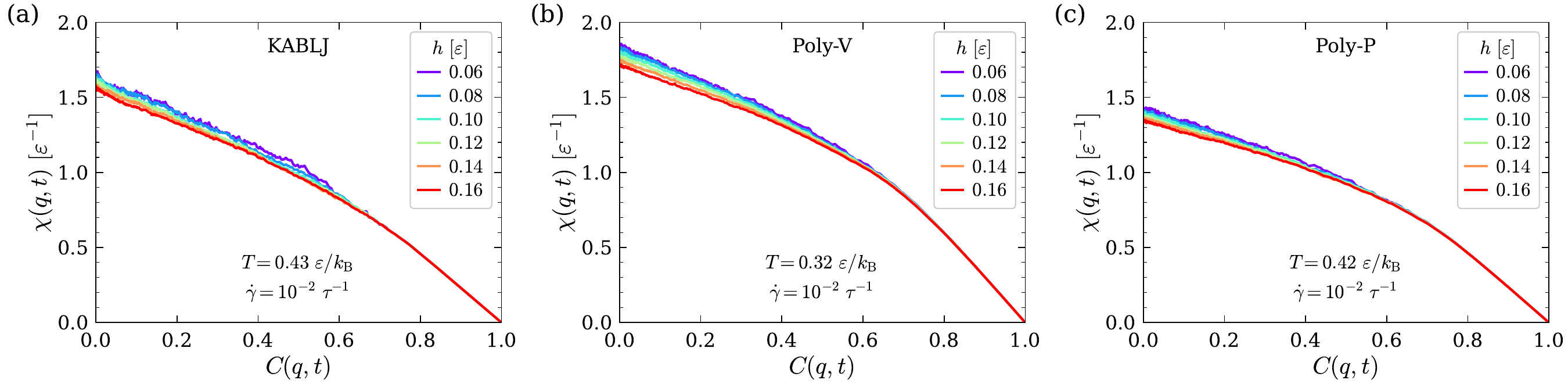}
	\caption{Influence of the perturbation strength $h$ on the fluctuation-dissipation relation (FDR). (a--c) Susceptibility $\chi(q, t)$ versus the correlation function $C(q, t)$ over a range of $h$ at a fixed $T$ indicated for the three systems, respectively. The shear rate is fixed at $\dot{\gamma} = 10^{-2} \, \tau^{-1}$. Here, our simulation data are obtained from $504$ independent runs.}
	\label{S4}
\end{figure}

\begin{figure}[htbp!]
	\centering
	\includegraphics[angle=0, width=0.6\textwidth]{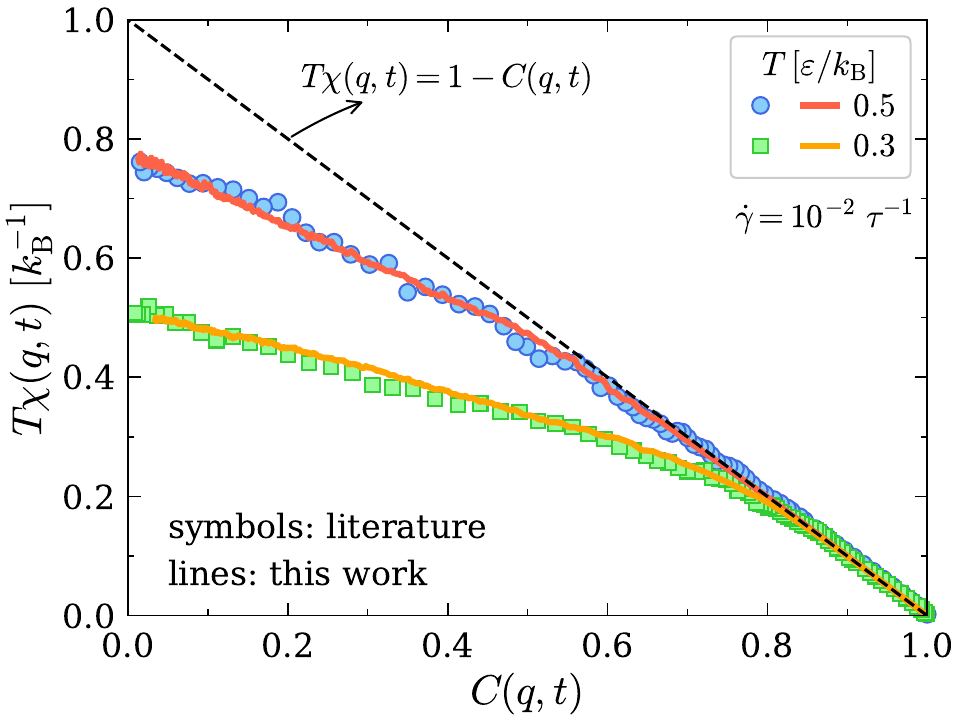}
	\caption{Validation of the numerical calculation of the FDR. The plot shows $T\chi(q, t)$ versus $C(q, t)$ at $T = 0.5\, \varepsilon / k_{\mathrm{B}}$ and $0.3 \,\varepsilon / k_{\mathrm{B}}$ for the KABLJ system. The shear rate is fixed at $\dot{\gamma} = 10^{-2} \, \tau^{-1}$. Symbols are data taken from Ref. \cite{Fluctuation_2000_63_012503} under the same conditions. The dash-dotted line shows the fluctuation-dissipation theorem (FDT) consistent with the bath temperature, as indicated. Here, our simulation data are obtained from $448$ independent runs. Note that the wave number is chosen to be $q = 7.47 \, \sigma^{-1}$, consistent with that in Ref. \cite{Fluctuation_2000_63_012503}.}
	\label{S5}
\end{figure}

\begin{figure}[htbp!]
	\centering
	\includegraphics[angle=0, width=0.975\textwidth]{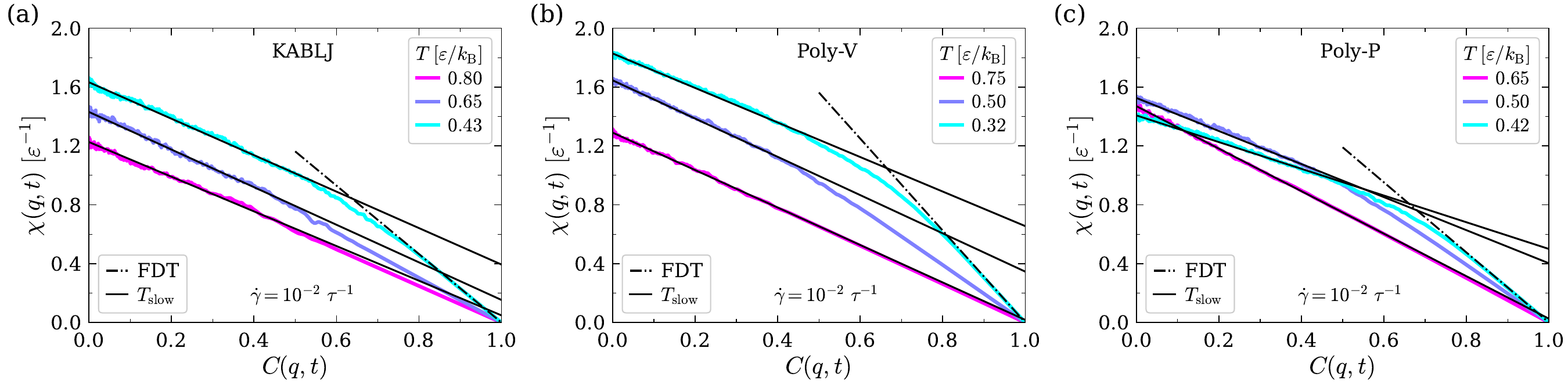}
	\caption{Temperature dependence of the FDR. (a--c) show $\chi(q, t)$ versus $C(q, t)$ at varying $T$ for the three systems, respectively. The shear rate is fixed at $\dot{\gamma} = 10^{-2} \, \tau^{-1}$. Dash-dotted lines show the FDT consistent with the bath temperature. Solid lines indicate the identification of an effective temperature $T_{\chi, \mathrm{slope}}$ associated with the slow modes of the sheared fluid. Here, our simulation data are obtained from $504$ independent runs.}
\label{S6}
\end{figure}

In our simulations, the system is driven at a constant shear rate $\dot{\gamma}$. Particular care is required in choosing the perturbation strength $h$ used to compute the FDR. A series of benchmark calculations indicates that $h=0.08\,\varepsilon$ provides a suitable compromise between statistical accuracy and linear-response behavior (Fig.~\ref{S4}), where $A_{\mathrm{ref}}$ in Eq.~\ref{Eq_chi_approx} denotes the unperturbed density fluctuation. This value is also consistent with the recommended range $h=0.05$--$0.2\,\varepsilon$ reported in Ref.~\cite{Fluctuation_2000_63_012503}. To improve statistics, all results are averaged over more than $500$ independent simulations.

To validate our implementation, Fig.~\ref{S5} compares our results with those reported in Ref.~\cite{Fluctuation_2000_63_012503} for the KABLJ system at $T=0.5\,\varepsilon/k_{\mathrm B}$ and $0.3\,\varepsilon/k_{\mathrm B}$ under a shear rate of $\dot{\gamma}=10^{-2} \, \tau^{-1}$. The agreement is essentially perfect. Figure~\ref{S6} further illustrates the temperature dependence of the FDR for all three systems. We find that $T_{\mathrm{slow}}$ closely follows the bath temperature $T$ at high temperatures but deviates significantly from it upon cooling. Notably, the FDR exhibits non-monotonic behavior in the Poly-P system, suggesting that the standard perturbation protocol, which neglects density fluctuations and their coupling to energy fluctuations, becomes problematic under isobaric conditions.

\subsection{Comparison of Effective Temperatures from Different Methods}
\label{Sec_Steady_ISF}

\begin{figure}[htbp!]
	\centering
	\includegraphics[angle=0, width=0.975\textwidth]{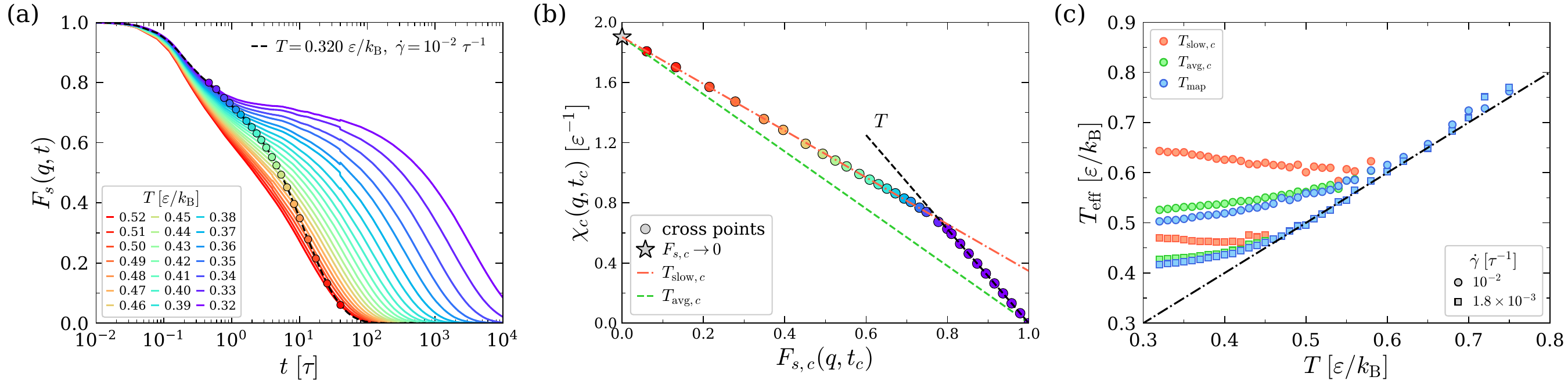}
	\caption{Determination of the effective temperature from the fluctuation-dissipation relation in the Poly-V system under steady shear. (a) Self-intermediate scattering function $F_s(q,t,T,\dot{\gamma})$ at $T=0.32\,\varepsilon/k_{\mathrm{B}}$ and $\dot{\gamma}=10^{-2}\,\tau^{-1}$, compared with equilibrium results at different temperatures $T$. Circles mark the intersections between the nonequilibrium and equilibrium curves used to define the crossing points at $t\equiv t_c$. (b) Parametric plot $\chi_c(q,t_c)$ versus $F_{s,c}(q,t_c)$ constructed from the intersection points shown in (a). The long-time branch slope defines $T_{\mathrm{slow},c}$, whereas the vertical-axis intercept $\chi_{c,\infty}=\lim_{F_{s,c}\to0}\chi_c$ gives $T_{\mathrm{avg},c}=1/(k_{\mathrm B}\chi_{c,\infty})$. (c) Effective temperatures $T_{\mathrm{map}}$, $T_{\mathrm{slow},c}$ and $T_{\mathrm{avg},c}$ as a function of the bath temperature $T$. The dash-dotted line indicates the equilibrium relation $T_{\mathrm{eff}}=T$.}
	\label{S14}
\end{figure}

\begin{figure}[htbp!]
	\centering
	\includegraphics[angle=0, width=0.975\textwidth]{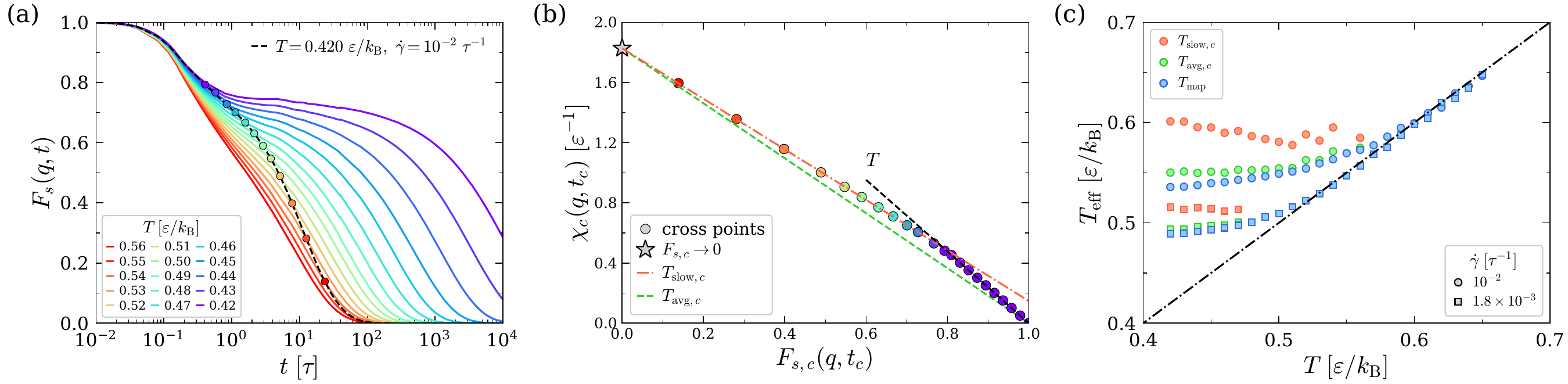}
	\caption{Determination of the effective temperature from the fluctuation-dissipation relation in the Poly-P system under steady shear. (a) Self-intermediate scattering function $F_s(q,t,T,\dot{\gamma})$ at $T=0.42\,\varepsilon/k_{\mathrm{B}}$ and $\dot{\gamma}=10^{-2}\,\tau^{-1}$, compared with equilibrium results at different temperatures $T$. Circles mark the intersections between the nonequilibrium and equilibrium curves used to define the crossing points at $t\equiv t_c$. (b) Parametric plot $\chi_c(q,t_c)$ versus $F_{s,c}(q,t_c)$ constructed from the intersection points shown in (a). The long-time branch slope defines $T_{\mathrm{slow},c}$, whereas the vertical-axis intercept $\chi_{c,\infty}=\lim_{F_{s,c}\to0}\chi_c$ gives $T_{\mathrm{avg},c}=1/(k_{\mathrm B}\chi_{c,\infty})$. (c) Effective temperatures $T_{\mathrm{map}}$, $T_{\mathrm{slow},c}$ and $T_{\mathrm{avg},c}$ as a function of the bath temperature $T$. The dash-dotted line indicates the equilibrium relation $T_{\mathrm{eff}}=T$.}
	\label{S15}
\end{figure}

\begin{figure}[htbp!]
 	\centering
 	\includegraphics[angle=0, width=0.975\textwidth]{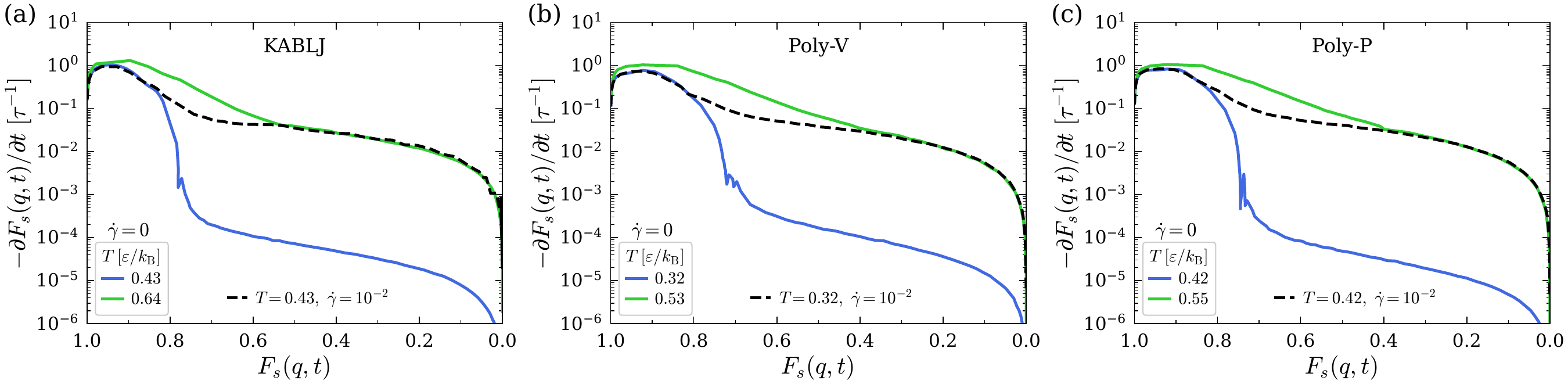}
 	\caption{Decay rate of the self-intermediate scattering function as a function of $F_s$. (a–c) show the results for the three systems at a low temperature with shear rate $\dot{\gamma}=10^{-2} \, \tau^{-1}$. As $F_s$ decreases beyond the cage regime, the sheared dynamics approach the equilibrium decay rate at $T_{\mathrm{avg},c}$, supporting both its long-time role and the asymptotic temperature selected by the $F_s\to0$ extrapolation.}
	\label{S8}
\end{figure}

\begin{figure}[htbp!]
	\centering
	\includegraphics[angle=0, width=0.975\textwidth]{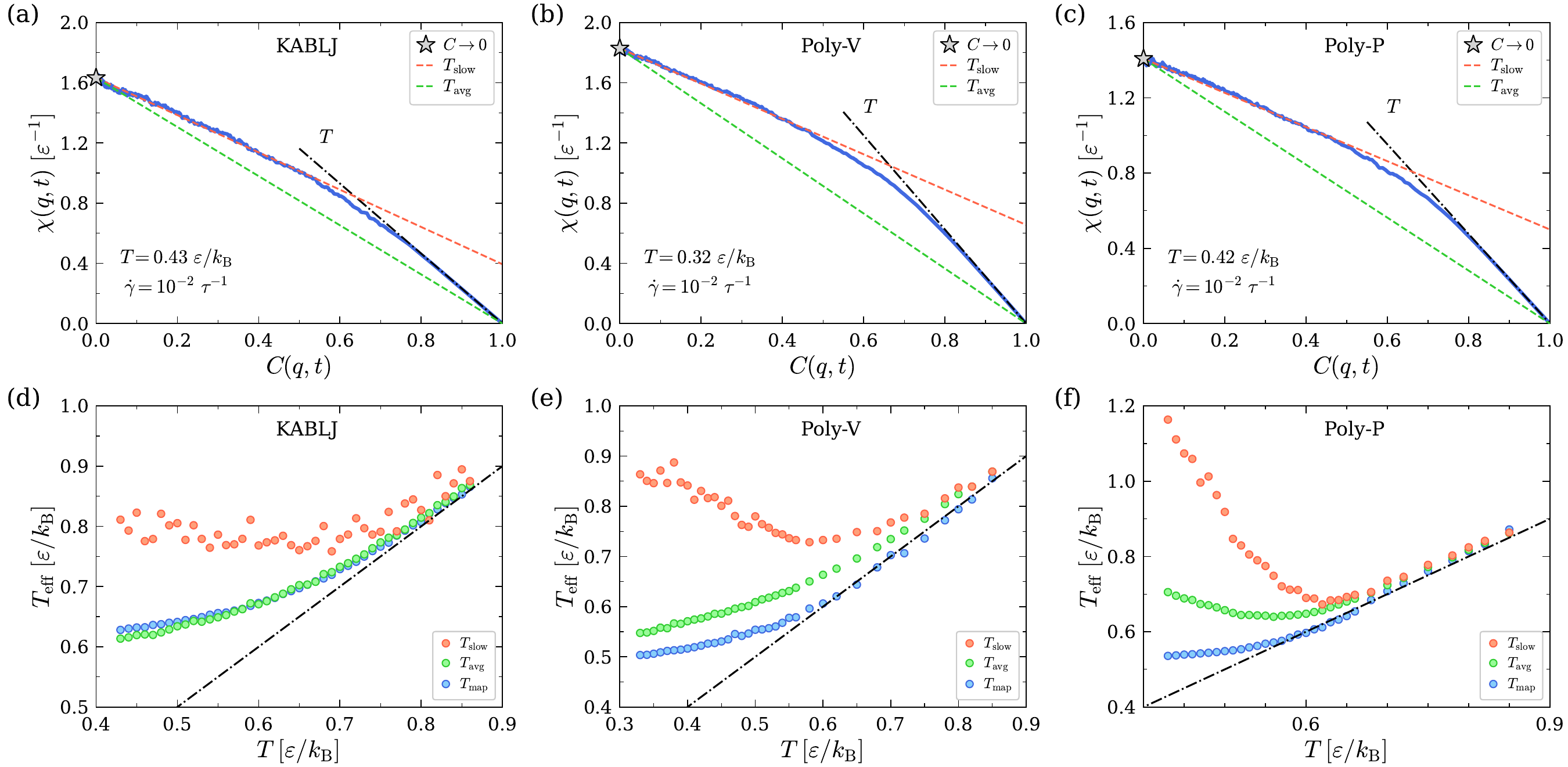}
	\caption{Determination of the effective temperature from the FDR in the three systems under steady shear. (a--c) Susceptibility $\chi(q,t)$ versus correlation $C(q,t)$ at $\dot{\gamma} = 10^{-2} \ \tau^{-1}$ for KABLJ ($T = 0.43\,\varepsilon/k_{\mathrm B}$), Poly-V ($T = 0.32\,\varepsilon/k_{\mathrm B}$), and Poly-P ($T = 0.42\,\varepsilon/k_{\mathrm B}$) systems. The long-time branch slope defines $T_{\mathrm{slow}}$, whereas the vertical-axis intercept $\chi_{\infty}=\lim_{C\to0}\chi$ gives $T_{\mathrm{avg}}=1/(k_{\mathrm B}\chi_{\infty})$. (d--f) Effective temperatures ($T_{\mathrm{map}}$, $T_{\mathrm{slow}}$, and $T_{\mathrm{avg}}$) as a function of the bath temperature $T$ for the corresponding systems. The dash-dotted line represents the equilibrium relation $T_{\mathrm{eff}} = T$.}
	\label{S7}
\end{figure}

In the crossing-point construction, we identify effective temperatures from the self-intermediate scattering function. By linearly extrapolating the long-time portion of the parametric curve to $F_{s,c}=0$, we extract the vertical-axis intercept $\chi_{c,\infty}=\lim_{F_{s,c}\to0}\chi_c(q,t_c)$. The average effective temperature is then defined as $T_{\mathrm{avg},c}=1/(k_{\mathrm B}\chi_{c,\infty})$. If expressed in terms of a slope, this definition uses the endpoint chord joining $(F_{s,c},\chi_c)=(1,0)$ and $(0,\chi_{c,\infty})$, whose slope is $-\chi_{c,\infty}$, while the slope of the long-time branch defines $T_{\mathrm{slow},c}$. For completeness, Figs.~\ref{S14} and~\ref{S15} show the the results associated with the FDR analysis based on $F_s(q,t)$ for the polymer systems.

Figure~\ref{S8} compares the decay rate, $-\partial F_s(q,t)/\partial t$, of each sheared system with its equilibrium counterparts. As $F_s$ decreases, the sheared decay rate progressively approaches the equilibrium rate at $T_{\mathrm{avg},c}$. This convergence provides independent dynamical evidence that $T_{\mathrm{avg},c}$ governs long-time structural relaxation and supports the temperature selected by extrapolating the crossing-point construction to $F_{s,c}=0$.

We now directly compare $T_{\mathrm{map}}$ with $T_{\mathrm{slow}}$ and $T_{\mathrm{avg}}$ determined from the FDR based on the perturbative method in Fig.~\ref{S7}. We find that neither $T_{\mathrm{slow}}$ nor $T_{\mathrm{avg}}$ follows $T_{\mathrm{map}}$ as the bath temperature $T$ is varied. In particular, $T_{\mathrm{slow}}$ initially follows $T$ at high $T$ but increases upon further lowering $T$, indicating that the standard FDR analysis does not provide a sensible effective temperature. We therefore conclude that the effective temperature defined through the average-mapping approach offers a more robust characterization of the nonequilibrium dynamics considered in our work.

\subsection{Comparison of Time Correlation Functions under Steady Shear and at Equilibrium}
\label{Sec_Steady_Comparison}

\begin{figure}[htbp!]
 	\centering
 	\includegraphics[angle=0, width=0.9\textwidth]{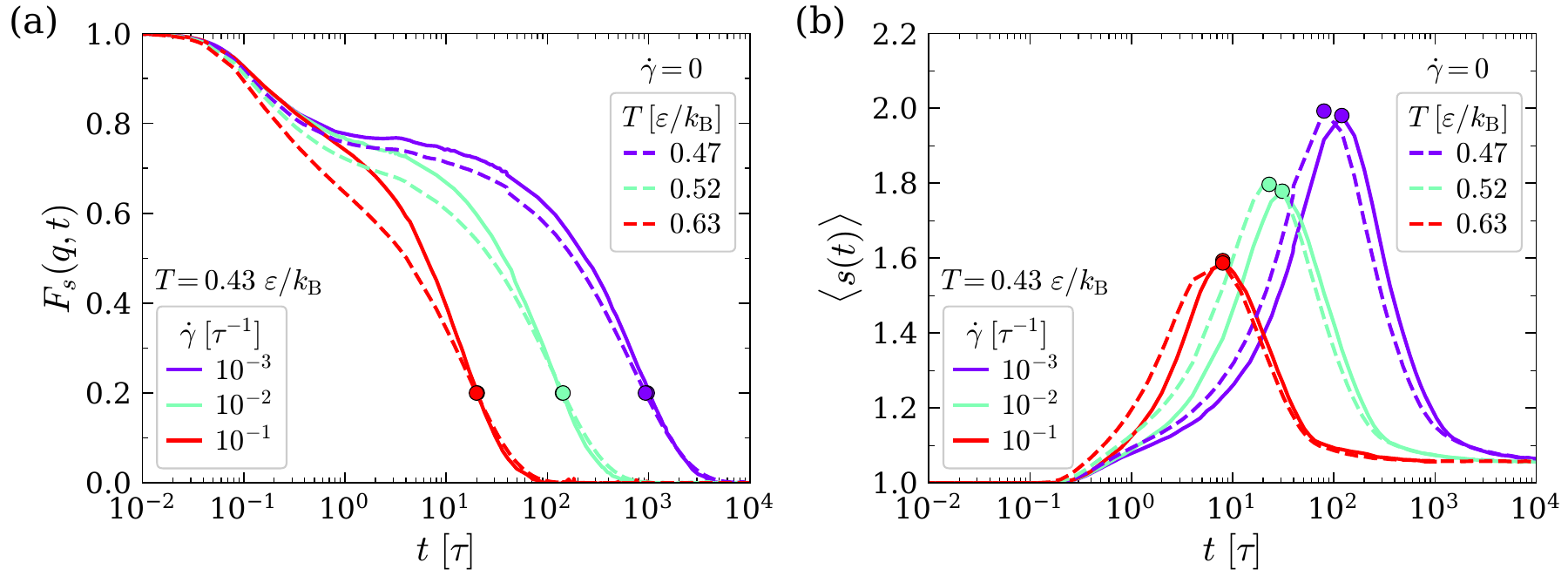}
 	\caption{Comparison of time correlation functions under steady shear and at equilibrium in the KABLJ system. (a,b) Self-intermediate scattering function $F_s(q,t)$ and number-averaged string length $\langle s(t)\rangle$ for the sheared fluid at $T=0.43\,\varepsilon/k_{\mathrm B}$ and different shear rates $\dot{\gamma}$, together with the corresponding equilibrium results at $T_{\mathrm{map}}$. Circles in (a) and (b) indicate the positions of $\tau_{\alpha}$ and $L$, respectively. Similar results are obtained for the Poly-V and Poly-P systems.}
 	\label{S9}
\end{figure}

In the main text, we adopt the definition of an effective temperature $T_{\mathrm{map}}$ obtained from the mapping of $\tau_{\alpha}$ indicated in Fig.~3(b). Here, we directly compare the corresponding time-correlation functions under steady shear and at equilibrium. Figure~\ref{S9} shows $F_s(q,t)$ and $\langle s(t)\rangle$ for the sheared fluid at $T=0.43\,\varepsilon/k_{\mathrm B}$ and different shear rates $\dot{\gamma}$, together with the corresponding equilibrium results at $T_{\mathrm{map}}$, where $T_{\mathrm{map}}$ is chosen as the closest available temperature in our simulations.

Although the relaxation time $\tau_{\alpha}$ of the sheared fluid at a given $\dot{\gamma}$ closely matches that of the equilibrium fluid at the corresponding $T_{\mathrm{map}}$, the full time dependence of $F_s(q,t)$ remains noticeably different [Fig.~\ref{S9}(a)]. A similar observation applies to $L$ and $\langle s(t)\rangle$ in Fig.~\ref{S9}(b). These results confirm that the concept of an effective temperature depends on the specific time and length scales used to characterize the dynamics, as discussed in the main text.

\section*{Code availability}

The code and scripts used to compute the fluctuation-dissipation relation are available on GitHub: \url{https://github.com/Zilong-Wang-2000/String-Model-Shear}.

\bibliographystyle{apsrev4-2}
\bibliography{refs}